\documentclass[10pt,onecolumn,journal]{IEEEtran}
\usepackage{url,amsmath,graphicx}
\usepackage{amsfonts}
\usepackage{multirow}
\usepackage[T1]{fontenc}
\usepackage[english]{babel}
\usepackage{bm}
\usepackage{textcomp}
\usepackage{float}
\usepackage{pifont}
\usepackage[left=2.5cm,top=2cm,right=2.5cm,bottom=2cm]{geometry}

\DeclareMathOperator{\argmin}{argmin}

\usepackage{amsfonts}
\usepackage{color}
\usepackage{multirow}
\usepackage{amssymb}
\usepackage{subfigure}
\usepackage[T1]{fontenc}
\usepackage{multirow}

\begin{document}
\title{Maximum Likelihood SNR Estimation of Linearly-Modulated Signals over Time-Varying   Flat-Fading SIMO Channels\vspace{-0.2cm} } 

\author{Faouzi Bellili, Rabii Meftehi, Sofi\`ene Affes, and Alex St\'{e}phenne
 \\\small INRS-EMT, 800, de la Gaucheti\`ere Ouest, Bureau 6900, Montreal, Qc, H5A 1K6, Canada.
\\\small Emails: \{bellili, meftehi, affes\}@emt.inrs.ca and  stephenne@ieee.org
\thanks{This work was accepted for publication in IEEE Transactions on signal Processing. Copyright (c) 2014 IEEE. Personal use of this material is permitted. However, permission to use this material for any other purposes must be obtained from the IEEE by sending a request to pubs-permissions@ieee.org. Work supported by a Canada Research Chair in Wireless Communications and by the Discovery Grants Program of NSERC. Work accepted for publication, in part, in IEEE ICASSP 2014. 
\cite{asilomar}.}}
\maketitle
\pagenumbering{arabic}\vspace {-2cm} 
\begin{abstract}
 In this paper, we tackle for the first time the problem of maximum likelihood (ML) estimation of the signal-to-noise ratio (SNR) parameter over time-varying single-input multiple-output (SIMO) channels. Both the data-aided (DA) and the non-data-aided (NDA) schemes are investigated. Unlike classical techniques where the channel is assumed to be slowly time-varying and, therefore, considered as constant over the entire observation period, we address the more challenging problem of \textit{instantaneous} (i.e., short-term or local)  SNR estimation over fast time-varying channels. The channel variations are tracked locally using a polynomial-in-time expansion. First, we derive in closed-form expressions the DA ML estimator and its bias. The latter is subsequently subtracted in order to obtain a new unbiased DA estimator whose  variance  and the corresponding  Cram\' er-Rao lower bound (CRLB) are also derived in closed form.  Due to the extreme nonlinearity of the log-likelihood function (LLF) in the NDA case, we resort to the expectation-maximization (EM) technique to iteratively  obtain the exact NDA ML SNR estimates within very few iterations. Most remarkably, the new EM-based NDA estimator is applicable to any linearly-modulated signal and  provides sufficiently accurate soft estimates (i.e., \textit{soft detection}) for each of the unknown transmitted symbols. Therefore,  \textit{hard detection} can be easily embedded in the iteration loop in order to improve its performance at low to moderate SNR levels. We show by extensive computer simulations that the new estimators are able to accurately estimate the \textit{instantaneous} per-antenna SNRs as they coincide with the DA CRLB over a wide range of practical SNRs. Moreover, the new EM-based NDA ML solution exhibits  substantial performance improvements against the SIMO-extended version of the estimator developed by Wiesel et al, referred to hereafter as WGM,  the only  benchmark  of the same class (i.e., NDA ML) suitable for proper comparisons.
\end{abstract}
\section{Introduction}
Over the recent years, there has been an increasing demand for the \textit{a priori} knowledge of the propagation environment conditions, fueled by an increasing thirst for taking advantage of any optimization opportunity that would enhance the system capacity. In essence, almost all the necessary information about these propagation conditions can be captured by estimating various channel parameters. In particular, the SNR is considered to be a key parameter whose \textit{a priori} knowledge can be exploited at both the receiver and the transmitter (through feedback), in order to reach the desired enhanced/optimal performance using various adaptive schemes. As examples, just to name a few, the SNR is required in all power control strategies, adaptive modulation and coding, turbo decoding, and handoff schemes [\ref{ref1}-\ref{ref3}]. 
SNR estimators can be broadly divided into two major categories: i) data-aided (DA) techniques in which the estimation process relies on a perfectly known (pilot) transmitted sequence, and ii) non-data-aided (NDA) techniques where the estimation process is applied with no \textit{a priori} knowledge about the transmitted symbols (but possibly the transmit constellation).\\
 DA approaches often provide sufficiently accurate estimates for  constant or quasi-constant parameters, even by using a reduced number of pilot symbols.
However, in fast changing wireless channels, they require larger pilot sequences in order to track the time variations of the unknown parameter. Indeed, when estimating the (time-varying) \textit{instantaneous} SNR from far-apart inserted  pilot symbols, the DA approaches are  unable to reflect the actual channel quality. This is because the receiver cannot accurately capture the details of the channel  between the pilot positions. In principle, this problem can be dealt with by inserting more pilot symbols. Unfortunately, this remedy results in an excessive overhead that entails  severe losses in system capacity. To circumvent this problem, NDA approaches are often considered instead for their ability to exploit both pilot and non-pilot received samples to estimate the channel coefficients. Consequently, they can provide the receiver with more refined channel tracking capabilities without impinging on the whole throughput of the system.\\
Historically, the problem of SNR estimation was first formulated and tackled in the context of single-input single-output (SISO) systems under \textit{constant} channels [\ref{ref18}, \ref{ref19}]. These two early estimators,  the well-known M2M4 technique among them, are moment-based ones. During the last decade, there has been a surge of interest in investigating this problem more intensively  and many  estimators tailored toward \textit{constant} SISO channels were introduced  [\ref{ref4}-\ref{Gappmair_EM_SISO}].
More recently, SNR estimation has also been  addressed under different types of diversity. In particular, a moment-based SNR estimator that exploits the across-antennae fourth-order  moments in \textit{constant} SIMO channels (i.e., spatial diversity) was proposed in [\ref{ref8}, \ref{ref22}].  ML SNR estimation has also been investigated in [\ref{boujelbene_conf}, \ref{ref10}] and [\ref{ref11}] under \textit{constant} SIMO and MIMO channels, respectively. Yet, current and future generation multi-antennae systems such as long-term-evolution (LTE), LTE-Advanced (LTE-A) and beyond (LTE-B) are expected to support reliable communications at very high velocities reaching $500$ Km/h [\ref{ref13}]. For such systems, classical assumptions of \textit{constant} channels  no longer hold and consequently  all the aforementioned SNR estimators shall  suffer from severe performance loss. Therefore, one needs to explicitly incorporate the channel time-variations in the estimation process and, so far, very few works have been  reported on this subject. In fact,   ML SNR estimation under SISO \textit{time-varying} channels was investigated in  [\ref{Morelli}, \ref{Abeida}] and \cite{ref14} for the DA and NDA modes, respectively. Under SIMO \textit{time-varying} channels, however, the only work that is available from the open literature is based on a least-squares (LS) approach [\ref{Faouzi}, \ref{Faouz_QBSC}].\\ 
Motivated by all these facts, we tackle in this paper the problem of ML \textit{instantaneous} SNR estimation over \textit{time-varying} SIMO channels, for  both the  DA and NDA schemes. Our proposed method is based on a piece-wise polynomial-in-time approximation for the channel process with very few unknown coefficients.
In the DA scenario where the receiver has access to a pilot sequence from which the SNR is obtained, the ML estimator is derived in closed form. Whereas in the NDA case where the transmitted sequence is partially unknown and random, the LLF becomes very complicated  and its  maximization is analytically intractable. Therefore, we resort to a more elaborate solution using the EM concept \cite{EM} and we develop thereby an iterative technique that is able to converge within very few iterations (i.e., in the range of 10). We also solve the challenging problem of local convergence that is inherent to all iterative techniques. In fact,  we propose an appropriate initialization procedure that guarantees the convergence of the new EM-based estimator to the global maximum of the  LLF which is indeed \textit{multimodal} under complex time-varying channels (in contrast to real channels). Most interestingly, the new EM-based SNR estimator is applicable for linearly-modulated signals in general (i.e., PSK, PAM, or QAM) and provides sufficiently accurate estimates [i.e., \textit{soft detection} (SD)] for the unknown transmitted symbols. Therefore,  \textit{hard detection} (HD) can be easily embedded in the iterative loop to further improve its performance over the  low-SNR region. Moreover, we develop a bias-correction procedure that is applicable in both the DA and NDA cases and which allows, over a wide practical SNR range, the new estimators to coincide  with the DA CRLB. Simulation results show the distinct performance advantage offered by fully exploiting the antennae \textit{diversity} and \textit{gain}
 in terms of instantaneous SNR estimation. In particular,  the new NDA estimator (either with SD or HD) shows overly superior performance against the most recent NDA ML technique\footnote{It is worth mentioning here that the very first EM-based ML SNR estimator was developed in \cite{Wiesel_EM_SISO}, but for \textit{constant} channels.}  both in its original SISO version \cite{ref14} and even in its SIMO-extended version developed here to further exploit the antennae \textit{gain}.\\
The remainder of this paper is structured as follows. In section II, we introduce  the system model that will be used throughout the article. In section III, we derive in closed form the new DA estimator with its bias and variance along with the corresponding CRLB. In section IV, we develop the new NDA EM-based ML estimator along with its appropriate initialization procedure. In section V, we  present  and analyze the simulation results before drawing out some concluding remarks  in section VI.\\
We  mention beforehand  that some of the common notations are adopted in  this paper. Indeed, vectors and matrices are represented in lower- and upper-case bold fonts, respectively. Moreover, $\{.\}^T$ and $\{.\}^H$ denote the transpose and the Hermitian (transpose conjugate) operators, respectively. The operators $\Re\{.\}$ and $\Im\{.\}$ return, respectively, the real and imaginary parts of any complex scalar or vector whereas $\{.\}^*$ returns its conjugate.  We also use ${\bf 0}_{K\times L}$ to denote a $(K\times L)$  zero matrix  and ${\bf 0}_{L}$ whenever $K=L$. 
\section{System Model}
Consider a digital transmission of  a $M-$ary linearly-modulated signal over a SIMO  communication system  under time-varying flat-fading channels.
 Assuming an ideal receiver with perfect time synchronization, and after matched filtering, the sampled baseband received signal over the $i^{th}$ antenna element, for $i=1,2,\cdots, N_r$,  can be expressed as:
\begin{equation}\label{Eq.1}
y_i(t_n)=h_i(t_n)a(t_n)+w_i(t_n),~~n=1,2,\cdots,N
\end{equation}
where $\{t_n=nT_s\}_{n=1}^N$ is the $n^{th}$ discrete-time instant, $T_s$ is the sampling period which is equal to the symbol period, and $N$ is the size of the observation window. We denote by $a(t_n)$ the linearly-modulated (i.e., M-PSK, M-PAM or M-QAM) transmitted symbol, by $y_i(t_n)$ the corresponding received sample, and by $h_i(t_n)$ the time-varying \textit{complex} channel gain, over each $i^{th}$ antenna branch. Note here that any carrier frequency offset (CFO) that is due to the Doppler shift and/or any mismatch between the transmitter and receiver local oscillators is absorbed in the \textit{complex} channel coefficients.  The noise components, $w_i(t_n)$, assumed to be temporally white and uncorrelated between antenna elements, are realizations of  zero-mean complex circular Gaussian processes, with independent real and imaginary parts, each of variance $\sigma^2$ (i.e., with overall noise power $N_0=2\sigma^2$). 
We assume that the same noise power is experienced over all the antenna branches (i.e., \textit{uniform} noise).\\
 The narrowband model in (\ref{Eq.1}) is well justified in practice by its wide adoption in current and next-generation \textit{multicarrier} communication systems, such as LTE, LTE-A and LTE-B  systems.  In fact, it is well known that OFDM systems transform a multipath frequency-selective channel in the time domain into a frequency-flat (i.e., narrowband) channel over each subcarrier as modeled by (\ref{Eq.1}).  Actually, multicarrier technologies were primarily designed to combat the multipath effects in high-data-rate communications  by bringing back the per-carrier propagation channel to the simple flat-fading case [\ref{multicarrier1}, \ref{multicarrier2}]. Yet, even over traditional single-carrier systems, the narrowband model in (1) could still be valid in practice when the symbol duration is smaller than the delay spread of the channel.
As mentioned in section I, however, most of the available techniques are based on the assumption that the channels are constant during the observation period, i.e., $h_i(t_n)=h_i$ for $n=1,2,\cdots,N$. But since in most real-world situations this assumption does not hold, one must incorporate the channel time variations in the SNR estimation process. Actually, all real-life channels have an essentially finite number of degrees of freedom due to restrictions on time duration or bandwidth (i.e., bandlimited). Consequently, their time variations can be efficiently captured through $t-$power series models \cite{bello}.
In fact, owing to the well-known Taylor's theorem, the time-varying channel coefficients can be locally tracked through a polynomial-in-time expansion of order ($L-1$) as follows:
\begin{eqnarray}\label{Eq.2}
\!\!\!\!\!\!\!\!\!\!\!h_i(t_n)&\!\!\!\!=\!\!\!\!&\sum_{l=0}^{L-1}\!c_i^{(l)}t_n^l+R^{(i)}_L(n),~i=1, 2, \cdots, N_r
\end{eqnarray}
where $c_i^{(l)}$ is the $l^{th}$ coefficient of the channel polynomial approximation over the $i^{th}$ branch among $N_r$ receiving antennae. The term $R^{(i)}_L(n)$ refers to the remainder of the Taylor series expansion. This remainder can be driven to zero under mild conditions such as i) a sufficiently high approximation order $(L-1)$, or ii) a sufficiently small ratio $\bar{N}F_D/F_s$ where  $F_s=1/T_s$ is the sampling rate, $F_D$ is the maximum Doppler frequency shift, and $\bar{N}$ is the size of the local approximation window. Choosing a high approximation order (i.e., first condition) may result in numerical instabilities due to badly conditioned matrices (depending on the value of the sampling rate). The second condition, however, can be easily fulfilled by choosing small-size local approximation windows (i.e., by appropriately selecting $\bar{N}$). By doing so, the remainder $R^{(i)}_L(n)$ can be neglected thereby yielding the accurate approximation:
\begin{eqnarray}\label{Eq.3}
h_i(t_n)=\sum_{l=0}^{L-1}c_i^{(l)}t_n^l,~~~~i=1, 2, \cdots, N_r.
\end{eqnarray}
Given all the received samples $\{y_i(n)\}_{n=1}^{N}$, for $i=1,2,\cdots,N_r$, and the statistical noise model,  our goal is to continuously estimate the \textit{instantaneous}\footnote{By ``\textit{instantaneous}'' SNR, we mean the ``\textit{local}'' or ``\textit{short-term}'' SNR that can be estimated from short observation windows.} per-antenna SNRs which are defined for each \{$i^{th}\}_{i=1}^{N_r}$ as follows: 
\begin{eqnarray}
\rho_i&=&\frac{\sum_{n=1}^{N}\big |h_i(t_n)|^2|a(t_n)|^2}{N(2\sigma^2)}\\
&=&\frac{\sum_{n=1}^{N}\left(|a(t_n)|^2 \left|\sum_{l=0}^{L-1}c_i^{(l)}t_n^l\right| ^2\right)}{N(2\sigma^2)}.
\end{eqnarray}
Note here that we do not make any other assumption about the channel coefficients than being unknown and deterministic. Of course, they might be random in practice. However,  we want to avoid  any \textit{a priori} knowledge about the  statistical model  of the channel. The motivation behind this choice is twofold: i) the statistical models are after all theoretical ones and as such they may not reflect the true behavior of real-world channels, and ii) the fading conditions (for instance the presence/absence of a line-of-sight component) might change in real time as users move from one location to another. In light of the above reasons, the new estimator is hence well geared toward any type of fading, a quite precious degree of freedom in practice. It is worth mentioning, though,  that estimators that capitalize on the statistical model of the fading channel, including the correlation in time between adjacent approximation windows, will generally perform better than those who do not. Although this research path  sounds interesting, it falls beyond the scope of this paper and may be treated in a future work.\\
Besides, the main advantage of local tracking is its ability to capture the unpredictable time variations of the channel gains  using  very few coefficients. Thus, we split up the entire observation window (of size $N$) into multiple local \textit{approximation} windows of size $\bar{N}$ (where $N$ is an integer multiple of $\bar{N}$). Then, after acquiring all the locally-estimated polynomial coefficients $\{\widehat{c}_{i,k}^{(l)}\}_{k=1}^{N/\bar{N}}$, where $k$ is the index of each local approximation window, and after averaging the local estimates of the single-sided noise power\footnote{These are indeed multiple estimates of the same constant but unknown parameter $\sigma^2$.}, $\{\widehat{\sigma^2_k}\}_{k=1}^{N/\bar{N}}$, the estimated SNRs are ultimately obtained for $i=1,2,\cdots,N_r$ as follows:
\begin{equation}\label{Eq.6}
\widehat{\rho}_i=\frac{\sum_{k=1}^{N/\bar{N}}\sum_{n=1}^{\bar{N}}\left|\widehat{a}_k(t_n)\right|^2\left |\sum_{l=0}^{L-1}\widehat{c}_{i,k}^{(l)}t_n^l\right|^2}{N\left(\frac{\bar{N}}{N}\sum_{k=1}^{N/\bar{N}}2\widehat{\sigma^2_k}\right)}.
\end{equation}
where, in the NDA case, $\big\{\widehat{a}_k(t_n)\big\}_{n=1}^{\bar{N}}$ are estimates of the unknown transmitted symbols corresponding to each $k^{th}$ local approximation window. Indeed, it will be seen in Section IV  that our NDA estimator is able to demodulate the transmitted symbols for any linearly-modulated signal. In the DA case, however,  $\big\{\widehat{a}_k(t_n)\big\}_{n=1}^{\bar{N}}$ are equal to the known transmitted symbols, i.e., $\big\{\widehat{a}_k(t_n)=a_k(t_n)\big\}_{n=1}^{\bar{N}}$.
\section{Derivation of the DA ML SNR Estimator and the DA CRLB}
In this section, we begin by deriving in closed-form expression the DA ML estimator for the SNR over each antenna element. Then, we will derive its bias revealing thereby that the derived estimator is actually biased due to the neglected remainder of the Taylor's series and the use of short observation windows. This will afterward allow us to obtain an unbiased version of the DA estimator by removing this bias during the estimation process.  We will also derive the closed-form expressions for the corresponding variance and CRLB. 
\subsection{Formulation of the DA ML SNR estimator}
In most real-world applications, some known pilot symbols are usually inserted to perform different synchronization tasks. The DA ML estimator can thus rely on these pilot symbols to estimate the \textit{instantaneous} SNR or at least to give a head start for an iterative algorithm (as will be derived in section IV) by providing a good initial guess about all the unknown parameters. Assume, therefore, that $N'$ such pilot or known symbols (out of $N$ pilot and non-pilot symbols) are periodically transmitted every $T_s'=N_pT_s$ where $N_p\geq 1$ is an integer quantifying the normalized (by $T_s$) time period between any two consecutive pilot positions. Here, we denote the size of the local approximation windows as $\bar{N}_{\textrm{DA}}$ (we shall later use $\bar{N}=\bar{N}_{\textrm{NDA}}$ in the NDA case). To begin with, we consider each antenna element, $i$, and gather the corresponding received pilot samples within each  $k^{th}$ approximation window in a column vector ${\bf y}'^{(k)}_{i,{\textrm{DA}}}=[y_i^{(k)}(t'_1), y_i^{(k)}(t'_2),\cdots, y_i^{(k)}(t'_{\bar{N}^{'}_{\textrm{DA}}})]^T$, where $t_n'=n~T_s'$ for $n=1,2,\cdots,\bar{N}^{'}_{\textrm{DA}}$. Here, $\bar{N}^{'}_{\textrm{DA}}=\bar{N}_{\textrm{DA}}/N_p$ is the number of pilot symbols in each approximation window which covers $\bar{N}_{\textrm{DA}}$ pilot and non-pilot received samples. Note also that  $\bar{N}_{\textrm{DA}}$ is a design parameter that can always be freely chosen as an integer multiple of $N_p$ (see section V for more details about the appropriate choice of $\bar{N}_{\textrm{DA}}$). The channel coefficients at each pilot position, $t_n'$, are also obtained from (\ref{Eq.3}) as follows:
\begin{eqnarray}\label{Eq.7}
h_{i,k}(t_n')=\sum_{l=0}^{L-1}c_{i,k}^{(l)}t_n'^l,~~~~i=1, 2, \cdots, N_r.
\end{eqnarray} 
For mathematical convenience, we  define the following vectors: 
\begin{eqnarray}
\label{Eq.8_channel}{\bf h}'_{i,k}&=&[h_{i,k}(t'_1), h_{i,k}(t'_2), \cdots, h_{i,k}(t'_{\bar{N}^{'}_{\textrm{DA}}})]^T\\
\label{Eq.8_noise}{\bf w}'_{i,k}&=&[w_{i,k}(t'_1),w_{i,k}(t'_2),\cdots,w_{i,k}(t'_{\bar{N}^{'}_{\textrm{DA}}})]^T\\
\label{Eq.8}{\bf c}_{i,k}&=&[c_{i,k}^{(0)}, c_{i,k}^{(1)}, \cdots, c_{i,k}^{(L-1)}]^T.
\end{eqnarray}
Over the $i^{th}$ antenna branch and the local approximation window $k$, ${\bf h}'_{i,k}$ contains the  \textit{complex} channel coefficients at pilot positions only and ${\bf w}'_{i,k}$ is the corresponding noise vector. The vector ${\bf c}_{i,k}$ contains the  coefficients of the local polynomial expansion. Then, using (\ref{Eq.7}), we can rewrite the  channel approximation model in a more compact form as follows:
\begin{equation}\label{Eq.9}
{\bf h}'_{i,k}={\bf T}'{\bf c}_{i,k},~~~~i=1,2,\cdots, N_r,
\end{equation}
where 
\begin{eqnarray}\label{Eq.10}
{\bf T}'=\begin{pmatrix} 1 & t'_1 & \cdots &{t'_1}^{L-1}\\ 1 & t'_2 & \cdots & {t'_2}^{L-1}\\ \vdots & \vdots & \ddots & \vdots\\ 1 & t'_{\bar{N}^{'}_{\textrm{DA}}} & \cdots & {t'}_{\bar{N}^{'}_{\textrm{DA}}}^{L-1}\end{pmatrix}.
\end{eqnarray}
Note that ${\bf T}'$ is a Vandermonde matrix with linearly-independent columns. Consequently, it is  full-rank  meaning that the pseudo-inverse that will appear in the sequel is always well defined. We further define ${\bf A}'_k=\textrm{diag}\big\{a_k(t'_1), a_k(t'_2),\cdots, a_k(t'_{\bar{N}^{'}_{\textrm{DA}}})\big\}$ to be the $({\bar{N}^{'}_{\textrm{DA}}} \times {\bar{N}^{'}_{\textrm{DA}}})$ diagonal matrix that contains all the known  symbols transmitted within the $k^{th}$ approximation window. 
Then, we can rewrite the corresponding received samples (over each  antenna element $i$) in a ${\bar{N}^{'}_{\textrm{DA}}}$-dimensional column vector as follows:
\begin{equation}\label{Eq.12}
{\bf y}'^{(k)}_{i,\textrm{DA}}={\bf A}'_k{\bf T}' {\bf c}_{i,k} + {\bf w}'_{i,k}={\bf \Phi}'_k{\bf c}_{i,k}+{\bf w}'_{i,k},
\end{equation}
where ${\bf \Phi}'_k={\bf A}'_k{\bf T}'$ is a known $({\bar{N}'_{\textrm{DA}}} \times L)$ matrix. We further stack all these per-antenna local observation vectors, $\{{\bf y}'^{(k)}_{i,\textrm{DA}}\}_{i=1}^{N_r}$, one below another into a single vector ${\bf y}'^{(k)}_{\textrm{DA}}=[{\bf y}'^{(k)T}_{1,\textrm{DA}}~{\bf y}'^{(k)T}_{2,\textrm{DA}}~\cdots~{\bf y}'^{(k)T}_{N_r,\textrm{DA}}]^T$.  By doing so, all the space-time received samples corresponding to the $k^{th}$ approximation window  can be written in a more succinct vector/matrix form as follows:
\begin{equation}\label{Eq.13}
{\bf y}'^{(k)}_{\textrm{DA}}={\bf B}'_k{\bf c}_k + {\bf w}'_k,
\end{equation}
where ${\bf c}_k=[{\bf c}_{1,k}^T~{\bf c}_{2,k}^T~\cdots~{\bf c}_{N_r,k}^T]^T$ and ${\bf w}'_k=[{{\bf w}'^T_{1,k}}~{{\bf w}'^T_{2,k}}~\cdots~{{\bf w}'^T_{N_r,k}}]^T$ are, respectively, $LN_r$- and ${\bar{N}'_{\textrm{DA}}}N_r$-dimensional column vectors vectorized in the same way and ${\bf B}'_k= \textrm{blkdiag}\{{\bf \Phi}'_k, {\bf \Phi}'_k, ... ,{\bf \Phi}'_k\}$ is a $({\bar{N}^{'}_{\textrm{DA}}}N_r \times LN_r)$ block-diagonal matrix.
The model in (\ref{Eq.13}) is a well-known linear model  in estimation theory for which the ML estimator along with its bias and variance  can be derived in closed form \cite{kay-estimation}. In fact, the probability density function (pdf) of the locally-observed vectors, ${\bf y}'^{(k)}_{\textrm{DA}}$, conditioned on ${\bf B}'_k$ and parameterized by ${\bm \theta}_k=[{\bf c}_k^T, \sigma^2]^T$ (a vector that contains all the unknown parameters over the $k^{th}$ approximation window) is given by:
\begin{eqnarray}\label{Eq.14}
\!\!\!\!\!\!\!\!\!\!\!\!p({\bf y}_{\textrm{DA}}^{(k)};{\bm \theta}_k\big |{\bf B}_k)&\!\!\!\!=\!\!\!\!&\frac{1}{(2\pi\sigma^2)^{{\bar{N}'_{\textrm{DA}}}N_r}}\times\nonumber\\
\!\!\!\!&\!\!\!\!\!\!\!\!&\!\!\!\!\!\!\!\!\!\!\!\!\!\!\!\!\exp\left\{-\frac{1}{2\sigma^2}[{\bf y}_{\textrm{DA}}^{(k)}-{\bf B}_k{\bf c}_k]^H[{\bf y}_{\textrm{DA}}^{(k)}-{\bf B}_k{\bf c}_k]\right\}.
\end{eqnarray}
The natural logarithm of (\ref{Eq.14}) yields the  DA LLF of the system as follows:
\begin{eqnarray}\label{Eq.15}
\!\!\!\!\!\!\!\!\!\!\!\!\!L_{\textrm{DA}}({\bm\theta}_k)&\!\!\!\!=\!\!\!\!&-{\bar{N}^{'}_{\textrm{DA}}}N_r\ln(2\pi)-{\bar{N}^{'}_{\textrm{DA}}}N_r\ln(\sigma^2)-\nonumber\\
\!\!\!\!\!\!\!\!\!\!\!\!\!\!\!\!\!\!\!\!&\!\!\!\!\!\!\!\!&~~~~~~~~~~~\frac{1}{2\sigma^2}[{\bf y}_{\textrm{DA}}^{(k)}-{\bf B}_k {\bf c}_k]^H[{\bf y}_{\textrm{DA}}^{(k)}-{\bf B}_k{\bf c}_k].
\end{eqnarray}
By differentiating (\ref{Eq.15}) with respect to the vector ${\bf c}_k$ and setting the result to zero, we obtain the ML estimate of the local polynomial coefficients over all the receiving antenna branches as follows:
\begin{equation}\label{Eq.16}
{\bf \widehat{c}}_{k,\textrm{DA}}=\left({{\bf B}'_k}^H{\bf B}'_k\right)^{-1}{{\bf B}'_k}^H{\bf y}'^{(k)}_{\textrm{DA}},
\end{equation}
where ${\bf T}'$ and ${\bf A}'_k$ are known matrices, and so is ${\bf B}'_k$ consequently. This is also the well-known least squares (LS) estimator which coincides with the ML estimator due to the   linearity of the observation model (\ref{Eq.13}) and the Gaussianity of the noise \cite{kay-estimation}. Note also that ${{\bf B}'_k}^H{\bf B}'_k$ is a block-diagonal matrix and thus its inverse can be easily obtained by computing the inverses of its small-size  diagonal blocks separately. To estimate the noise variance, we first find the partial derivative of (\ref{Eq.15}) with respect to $\sigma^2$. Then after setting it to zero and substituting ${\bf c}_k$ by ${\bf \widehat{c}}_{k,\textrm{DA}}$ obtained in (\ref{Eq.16}), the ML estimate for the noise variance is derived as follows:
\begin{eqnarray}\label{Eq.17}
\!\!\!\!\!\!\!\!\widehat{\sigma^2}_{k,\textrm{DA}}&\!\!\!\!=\!\!\!\!&\displaystyle\frac{1}{2{\bar{N}^{'}_{\textrm{DA}}}N_r}[{\bf y}_{\textrm{DA}}^{(k)}-{\bf B}_k {\bf \widehat{c}}_{k,\textrm{DA}}]^H[{\bf y}_{\textrm{DA}}^{(k)}-{\bf B}_k {\bf \widehat{c}}_{k,\textrm{DA}}].
\end{eqnarray}
Actually, combining (\ref{Eq.16}) and (\ref{Eq.17}), it can be further shown that:
\begin{eqnarray}
\widehat{\sigma^2}_{k,\textrm{DA}}&=&\frac{1}{2{\bar{N}^{'}_{\textrm{DA}}}N_r}\left[{{\bf y}'^{(k)}_{\textrm{DA}}}^H({\bf I}-{\bf P}_k){\bf y}'^{(k)}_{\textrm{DA}}\right],\nonumber\\
\label{projector_manuscript}&=&\frac{1}{2{\bar{N}^{'}_{\textrm{DA}}}N_r}\left[{{\bf y}'^{(k)}_{\textrm{DA}}}^H{\bf P}_k^{\perp}{\bf y}'^{(k)}_{\textrm{DA}}\right],
\end{eqnarray}
in which ${\bf P}_k={\bf B}'_k\left({{\bf B}'_k}^H{\bf B}'_k\right)^{-1}{{\bf B}'_k}^H$ and ${\bf P}^{\perp}_k={\bf I}-{\bf P}_k$ are, respectively, the projection matrices onto the column space of ${\bf B}'_k$ (i.e., signal subspace) and its orthogonal complement (i.e., noise subspace).  
In order to obtain the estimated SNRs over the entire observation window for a given $i^{th}$ antenna element, we begin by extracting the locally-estimated polynomial coefficients, $\{{\bf \widehat{c}}_{i,{\textrm{DA}}}^{(k)}\}_k$. Then  the channel coefficients\footnote{The DA SNR estimator is able to implicitly identify the time-varying channel coefficients and estimate the noise power. Yet study and assessment of these capabilities or functionalities fall beyond the scope of this paper.} corresponding to the pilot positions over each approximation window are obtained as $\{{\bf \widehat{h}}'^{(k)}_{i,\textrm{DA}}={\bf T}'{\bf \widehat{c}}_{i,{\textrm{DA}}}^{(k)}\}_k$. The latter are then  stacked into a single vector ${\bf \widehat{h}}'_{i,\textrm{DA}}=\left[{\bf \widehat{h}}'^{(1)}_{i,\textrm{DA}},{\bf \widehat{h}}'^{(2)}_{i,\textrm{DA}},\cdots,{\bf \widehat{h}}_{i,\textrm{DA}}'^{(N/\bar{N}_{\textrm{DA}})}\right]^T$. On the other hand, the local estimates for the noise variance are averaged over all the local approximation windows:
\begin{eqnarray}\label{noise_variance_DA}
\widehat{\sigma^2}_{\textrm{DA}}=\frac{\bar{N}_{\textrm{DA}}}{N}\sum_{k=1}^{N/\bar{N}_{\textrm{DA}}}\widehat{\sigma^2}_{k,\textrm{DA}},
\end{eqnarray}
to finally obtain the DA ML SNR estimator over each antenna element as follows:
\begin{equation}\label{Eq.18}
\widehat{\rho}_{i,\textrm{DA}}=\displaystyle\frac{\big|\big|\displaystyle\mathbf{A}'\widehat{\bf h}'_{i,\textrm{DA}}\big|\big|^2}{\frac{N}{N_p}(2\widehat{\sigma^2}_{\textrm{DA}})},~~~i=1,2,,\cdots,N_r
\end{equation}
with $\mathbf{A}'=\textrm{blkdiag}\left\{\mathbf{A}'_1,\mathbf{A}'_2\cdots,\mathbf{A}'_{N/\bar{N}_{\textrm{DA}}}\right\}$ being a known ($N/N_p\times N/N_p$) diagonal matrix that contains all the  pilot symbols transmitted over the whole observation window. 
\subsection{Derivation of the exact bias and variance for the DA ML SNR  estimator}
To improve the accuracy of the DA ML SNR  estimator, we calculate and remove its bias. After doing so, we will derive the exact expression for the variance of the resulting unbiased estimator. Here, for  reasons that shall become clear later in sections IV and V, we are interested in assessing the performance of the ``\textit{completely DA}'' estimator for which all the $N$ transmitted  symbols are assumed to be pilots, i.e., $\bar{N}^{'}_{\textrm{DA}}=\bar{N}_{\textrm{DA}}$  \big(or equivalently $N_p=1$ and hence $N'=N$\big). In a nutshell, our ultimate goal is to develop a bias-correction procedure that is also valid for the NDA estimator to be derived in the next section. As will be seen there, the NDA estimator is able to correctly demodulate all the transmitted symbols which can then be treated (all) as pilots by the receiver. Thus, the same bias-correction procedure developed hereafter can also be applied in order to obtain an  \textit{unbiased} version of the \textit{biased} NDA estimator.  To begin with, recall from (\ref{Eq.6}) that  the ML DA SNR estimates are given in the ``\textit{completely DA}'' scenario by:
\begin{equation}\label{Eq.19}
\widehat{\rho}_{i,\textrm{DA}}=\frac{\sum_{k=1}^{N/\bar{N}_{\textrm{DA}}}\sum_{n=1}^{\bar{N}_{\textrm{DA}}}\left(\left|a_k(t_n)\right|^2\left |\sum_{l=0}^{L-1}\widehat{c}_{i,k}^{(l)}t_n^l\right|^2\right)}{ N\left(\frac{\bar{N}_{\textrm{DA}}}{N}\sum_{k=1}^{N/\bar{N}_{\textrm{DA}}}2\widehat{\sigma^2}_{k,\textrm{DA}}\right)},
\end{equation}
from which we show in Appendix A the following theorem:\\
\indent\textbf{\textit{Theorem 1:}} the DA ML SNR estimator in (\ref{Eq.19}) is a scaled noncentral $F$ distributed random variable, i.e:
\begin{equation}\label{Eq.20}
\frac{(N-\frac{N}{\bar{N}_{\textrm{DA}}}L)}{\frac{N}{\bar{N}_{\textrm{DA}}}L}\widehat{\rho}_{i,\textrm{DA}}=F_{v_1,v_2}(\lambda),
\end{equation}
where $F_{v_1,v_2}(\lambda)$ is the noncentral $F$ distribution with a noncentrality parameter $\lambda=N\rho_i$ and degrees of freedom $v_1=\frac{N}{\bar{N}_{\textrm{DA}}}L$ and $v_2=N_r(N-\frac{N}{\bar{N}_{\textrm{DA}}}L)$.\\
\indent \textit{\textbf{Proof:}} see Appendix A.\\
Hence, the mean and the variance of the new DA ML SNR estimator follow immediately from the following two expressions:
\begin{equation}\label{Eq.21}
\textrm{E}\{F\}=\frac{v_2(v_1+\lambda)}{v_1(v_2-2)},~~~v_2>2,
\end{equation}
\begin{equation}\label{Eq.22}
\textrm{Var}\{F\}=2\left(\frac{v_2}{v_1}\right)^2\frac{(v_1+\lambda)^2+(v_1+2\lambda)(v_2-2)}{(v_2-2)^2(v_2-4)},~~~v_2>4.
\end{equation}
Indeed, using (\ref{Eq.20}) through (\ref{Eq.22}) and denoting $\epsilon= L/\bar{N}_{\textrm{DA}}$, we show in Appendix B the following two identities:
\begin{eqnarray}\label{Eq.23}
\textrm{E}\{\widehat{\rho}_{i,\textrm{DA}}\}=\displaystyle\frac{N_rN}{N_rN(1-\epsilon)-1}\left(\rho_i+\frac{\epsilon}{2}\right),
\end{eqnarray}
\begin{eqnarray}\label{Eq.24}
\textrm{Var}\{\widehat{\rho}_{i,\textrm{DA}}\}&\!\!\!=\!\!\!&\displaystyle\frac{(N_rN)^2\bigg[\displaystyle\rho_i^2+\rho_i\!\left(\!2N_r(1-\epsilon)+\epsilon-\frac{2}{N}\!\right)\!+\!\left(\!\frac{N_r}{2}-\frac{1}{2N}\!\right)\epsilon-\left(\!\frac{N_r}{2}-\frac{1}{4}\right)\epsilon^2\bigg]}{\bigg(N_rN(1-\epsilon)-1\bigg)^2\bigg(N_rN(1-\epsilon)-2\bigg)}.
\end{eqnarray}
\noindent Now, using (\ref{Eq.23}) we can derive the exact bias for the DA estimator as follows:
\begin{eqnarray}\label{Eq.25}
\textrm{Bias}\{\widehat{\rho}_{i,\textrm{DA}}\}&\!\!\!\!=\!\!\!\!&\rho_i\bigg(\frac{N_rN}{N_rN(1-\epsilon)-1}-1\bigg)+\frac{N_rN\epsilon}{2N_rN(1-\epsilon)-1}\nonumber,
\end{eqnarray}
which is not identically zero meaning that the estimator is  biased. Actually, this bias is in part due to the use of a limited number of received samples during the estimation process and in part due to dropping the Taylor's remainder in the channel approximation model. Yet, an unbiased version of this DA estimator \big(i.e., $\textrm{E}\{\widehat{\rho}_{i,\textrm{DA}}^{~\textrm{UB}}\}=\rho_i$\big) can  be straightforwardly obtained from (\ref{Eq.23}) as follows:
\begin{eqnarray}\label{Eq.26}
\widehat{\rho}_{i,\textrm{DA}}^{~\textrm{UB}}=\frac{N_rN(1-\epsilon)-1}{N_rN}\widehat{\rho}_{i,\textrm{DA}}-\frac{\epsilon}{2}.
\end{eqnarray}
Therefore, by combining (\ref{Eq.24}) and (\ref{Eq.26}), it follows that:
\begin{eqnarray}\label{Eq.28}
\textrm{Var}\{\widehat{\rho}_{i,\textrm{DA}}^{~\textrm{UB}}\}&\!\!\!\!=\!\!\!\!&\frac{1}{NN_r(1-\epsilon)-2} \bigg[\rho_i^2+\rho_i\!\left(\!2N_r(1-\epsilon)+\epsilon-\frac{2}{N}\!\right)\nonumber\\
&\!\!\!\!\!\!\!\!&~~~~~~~~~~~~~~~+\left(\!\frac{N_r}{2}-\frac{1}{2N}\!\right)\epsilon-\left(\!\frac{N_r}{2}-\frac{1}{4}\right)\epsilon^2\bigg].\nonumber
\end{eqnarray}
In practice, the variance of unbiased estimators is usually compared to the so-called Cram\'er-Rao lower bound (CRLB) which is a fundamental benchmark that reflects the best achievable performance ever. Therefore, as detailed in Appendix C, we also derive the CRLB for  DA  SNR estimation over time-varying channels as follows: \begin{eqnarray}\label{Eq.36}
\textrm{CRLB}_\textrm{DA}(\rho_i)=\frac{\rho_i}{N}\left(2+\frac{\rho_i}{N_r}\right).
\end{eqnarray}  
Now, by closely inspecting (\ref{Eq.28}),  it can be verified that the mean square error (or the variance) of the unbiased estimator $\textrm{MSE}\{\widehat{\rho}_{i,\textrm{DA}}^{~\textrm{UB}}\}=\textrm{E}\{\big(\widehat{\rho}_{i,\textrm{DA}}^{~\textrm{UB}}-\rho_i\big)^2\}$ tends  asymptotically\footnote{It should be mentioned here that the second asymptotic condition, $\bar{N}_{\textrm{DA}}\gg L$, must indeed be taken into account. This is because the estimates of the channel coefficients, over each approximation window, are obtained from the $\bar{N}_{\textrm{DA}}$ samples received over that window only.  Their accuracy does not depend, therefore, on how many samples are received outside the considered approximation window (the rest of the observation interval). Yet, the size of the whole observation window, $N$, will ultimately affect the performance of the SNR estimator through the noise variance estimate that is indeed obtained from all the received samples.}, i.e., when $N\gg 1$ and $\bar{N}_{\textrm{DA}}\gg L$ (or equivalently $\epsilon \ll 1$), to the aforementioned CRLB, i.e.:
\begin{eqnarray}\label{Eq.29}
\textrm{MSE}\{\widehat{\rho}_{i,\textrm{DA}}^{~\textrm{UB}}\}=\textrm{Var}\{\widehat{\rho}_{i,\textrm{DA}}^{~\textrm{UB}}\}\longrightarrow\textrm{CRLB}_\textrm{DA}(\rho_i),
\end{eqnarray}
 Therefore, our unbiased DA ML estimator is asymptotically efficient and attains the theoretical optimal  performance as will be validated by computer simulations in section V.
 In addition, even though the CRLB in (\ref{Eq.36}) was primarily derived for the DA scenario, it  will also  hold in the NDA case\footnote{Note here that the derivation of NDA CRLBs (especially the \textit{stochastic} ones) are extremely challenging in presence of linearly-modulated signals, in general, and that they usually deserve stand-alone contributions even in the very basic case of \textit{constant} SISO channels [\ref{bellili_CRLB_conf}, \ref{bellili_CRLB_journal}], \ref{bellili_CRLB_journal_CFO}} for moderate to high SNR values. This is hardly surprising since the NDA algorithm developed in the next section is able to perfectly estimate/detect all the unknown transmitted symbols over this SNR region,  reaching thereby the \textit{ideal} DA performance. In other words, the new NDA ML estimator derived next will be able to reach the performance achievable in ideal conditions (i.e., perfect knowledge about all the transmitted symbols).    
\section{Derivation of the new EM-based ML SNR estimator}
In this section, we derive the new NDA ML SNR estimator where  partial or no {\it a priori} knowledge about the transmitted symbols is assumed at the receiver.  The constellation type and order, however, are  assumed to be known to the receiver. 
\subsection{Formulation of the new NDA ML SNR estimator}
\noindent To begin with, we mention that the  problem formulation adopted in the DA case is problematic in the NDA scenario. In fact, as will be seen shortly, the EM algorithm averages the likelihood function, at each iteration, over all the possible values of the unknown transmitted symbols. Consequently, by adopting the same formulation of section III, the EM algorithm would average over all the possible realizations of the matrix {\bf B} that contains the whole transmitted sequence. This results in a combinatorial problem with prohibitive (i.e., exponentially increasing) complexity. Typically, its complexity would be of order $\mathcal{O}(M^N)$ where $M$ is the modulation order and $N$ is the size of the observation window. In the DA scenario, this was feasible since the matrix {\bf B} (or the transmitted sequence) is \textit{a priori} known to the receiver and no averaging was required. Thus, we reformulate our system differently so that the EM algorithm averages over the \textit{elementary} symbols transmitted at separate time instants instead of averaging over the whole transmitted sequence. In this way, the complexity of the algorithm becomes only linear with the modulation order and the observation window size.\\
To that end, we define\footnote{For the sake of simplifying notations  in what follows, we shall  use  ${\bf t}(n)$ instead of ${\bf t}(nT_s)$ and keep  dropping $T_s$ in all similar quantities.} the vector ${\bf t}(n)=[1, t_n,t_n^2, \cdots, t_n^{L-1}]^T$ which is the $n^{th}$ row (transposed to a column vector) of the Vandermonde time matrix, ${\bf T}_{\bar{N}_{\textrm{NDA}}}$, defined as:
\begin{eqnarray}\label{Eq.39}
{\bf T}_{\bar{N}_{\textrm{NDA}}}=\begin{pmatrix} 1 & t_1 & \cdots &{t_1}^{L-1}\\ 1 & t_2 & \cdots & {t_2}^{L-1}\\ \vdots & \vdots & \ddots & \vdots\\ 1 & t_{\bar{N}_{\textrm{NDA}}} & \cdots & {t}_{\bar{N}_{\textrm{NDA}}}^{L-1}\end{pmatrix},
\end{eqnarray}
and rewrite the channel model as follows:
\begin{eqnarray}\label{Eq.44}
h_{i,k}(t_n)&=&\sum_{l=0}^{L-1}c_{i,k}^{(l)}t_n^l~=~{\bf c}_{i,k}^T {\bf t}(n).
\end{eqnarray}
At each time instant $n$ (within the $k^{th}$ approximation window of size\footnote{Note that the local approximation windows in the DA and NDA scenarios might have different sizes $\bar{N}_{\textrm{DA}}$ and $\bar{N}_{\textrm{NDA}}$, respectively.} $\bar{N}=\bar{N}_{\textrm{NDA}}$), we stack all the received samples at the output of the antennae array, $\{y_{i,k}(n)\}_{i=1}^{N_r}$, known as \textit{snapshot} in array signal processing terminology, into a single vector, ${\bf y}_k(n)=[y_{1,k}(n), y_{2,k}(n),\cdots,y_{N_r,k}(n)]^T$, which can be expressed as:
\begin{equation}\label{Eq.45}
{\bf y}_k(n)=a_k(n){\bf C}_k{\bf t}(n)+{\bf w}_k(n),
\end{equation}
in which $a_k(n)$ is the corresponding unknown transmitted symbol, ${\bf C}_k=[{\bf c}_{1,k}, {\bf c}_{2,k},\cdots, {\bf c}_{N_r,k}]^T$ and ${\bf w}_k(n)=[w_{1,k}(n),w_{2,k}(n), ... ,w_{N_r,k}(n)]^T$. Note that the vectors ${\bf c}_{i,k}$ were defined previously in (\ref{Eq.8}). From (\ref{Eq.45}), the pdf of the received vector, ${\bf y}_k(n)$, conditioned on the transmitted symbol $a_k(n)$, can be expressed as the product of its element-wise pdfs as follows:
\begin{eqnarray}\label{Eq.46}
\!\!\!\!\!\!\!\!p\big({\bf y}_k(n);{\bm\theta}_k|a_k(n)=a_m\big)&\!\!\!\!=\!\!\!\!&\frac{1}{\left(2\pi\sigma^2\right)^{N_r}}\times\nonumber\\
\!\!\!\!\!\!\!\!&\!\!\!\!\!\!\!\!&\!\!\!\!\!\!\!\!\!\!\!\!\!\!\!\!\!\!\!\!\!\!\!\!\!\!\!\!\!\!\!\!\!\!\!\!\!\!\exp\left\{-\frac{1}{2\sigma^2}\sum_{i=1}^{N_r}\big|y_{i,k}(n)- a_m{\bf c}_{i,k}^T{\bf t}(n)\big|^2\right\},
\end{eqnarray}
in which $a_m$ is the \textit{hypothetically} transmitted symbol that is randomly drawn from the $M$-ary constellation alphabet $\mathcal{C}=\{a_1,a_2,\cdots,a_M\}$.
Now, averaging (\ref{Eq.46}) over this alphabet and assuming the transmitted symbols to be equally likely, i.e., $P[a_m]=1/M$ for $m=1,2,\cdots,M$, the pdf of the received vector is obtained as:
\begin{eqnarray}\label{Eq.47}
\!\!\!\!\!\!\!\!\!\!\!\!\!\!\!\!p\big({\bf y}_k(n);{\bm\theta}_k\big)&\!\!\!\!=\!\!\!\!&\frac{1}{M}\sum_{m=1}^{M}p\big({\bf y}_k(n);{\bm\theta}_k|a_k(n)=a_m\big)\nonumber\\
\!\!\!\!\!\!\!\!&\!\!\!\!\!\!\!\!&\!\!\!\!\!\!\!\!\!\!\!\!\!\!\!\!\!\!\!\!\!\!\!\!\!\!\!\!\!\!\!\!\!\!\!\!=\frac{\sum_{m=1}^{M}\exp\left\{-\frac{1}{2\sigma^2}\sum_{i=1}^{N_r}\big|y_{i,k}(n)- a_m{\bf c}_{i,k}^T{\bf t}(n)\big|^2\right\}}{M\left(2\pi\sigma^2\right)^{N_r}}.
\end{eqnarray}
By inspecting (\ref{Eq.47}), it becomes clear that a joint maximization of the likelihood function with respect to $\sigma^2$ and $\{{\bf c}_{i,k}\}^{N_r}_{i=1}$ is analytically intractable. Yet, this multidimensional optimization problem can be efficiently tackled using the EM concept after defining the right \textit{incomplete} and \textit{complete} data sets. In fact,  we define  at a per-snapshot basis (in array signal processing terminology) multiple ``\textit{incomplete}'' data sets  each of which containing the $N_r$ samples received  at a given time instant $nT_s$ [i.e., ${\bf y}_k(n)$].  Each of these ``\textit{incomplete}'' data sets is completed by the single unknown symbol, $a_k(n)$, corresponding to the same snapshot. Then, the LLF, $L({\bm\theta}_k|a_k(n)=a_m)\triangleq\ln\big(p({\bf y}_k(n);{\bm\theta}_k\big|a_k(n)=a_m)\big)$, of ${\bf y}_k(n)$ conditioned on the transmitted symbol $a_k(n)$  is given by:
\begin{eqnarray}\label{Eq.48}
L({\bm\theta}_k|a_k(n)=a_m)&\!\!\!\!=\!\!\!\!&-N_r\ln(2\pi\sigma^2)-\nonumber\\
&\!\!\!\!\!\!\!\!&~~~~~~~\frac{1}{2\sigma^2}\sum_{i=1}^{N_r}\big|y_{i,k}(n)-a_m{\bf c}_{i,k}^T{\bf t}(n)\big|^2\nonumber\\ 
&\!\!\!\!=\!\!\!\!&-N_r\ln(2\pi\sigma^2)\!-\!\frac{1}{2\sigma^2}\sum_{i=1}^{N_r}\bigg (\!|y_{i,k}(n)|^2+\nonumber\\
&\!\!\!\!\!\!\!\!&\!\!\!\!\!\!\!\!\!\!\!\!\!\!\!\!\!\!\!\!\!\!\!\!\!\!|a_m|^2\big|{\bf c}_{i,k}^T{\bf 
t}(n)\big|^2\!-\!2\Re\left\{y_{i,k}^{*}(n)a_m{\bf c}_{i,k}^T{\bf t}(n)\right\}\!\bigg).
\end{eqnarray}
The new EM-based algorithm runs in two main steps. During the ``expectation step'' (E-step), the expected value of the above likelihood function with respect to all the possible transmitted symbols $\{a_m\}_{m=1}^M$ is computed. Then, during the ``maximization-step'' (M-step), the output of the E-step is maximized with respect to  all the unknown parameters. The E-step is established as follows: starting from an initial guess\footnote{Initialization is  critical to the convergence of the new iterative NDA algorithm. It will be discussed in more details in section IV-B.}, $\widehat{{\bm\theta}_k}^{(0)}$, of the unknown parameter vector, the objective function is updated iteratively according to:
\begin{eqnarray}\label{Eq.49}
Q\!\left(\!{\bm\theta}_k|\widehat{{\bm \theta}_k}^{(q-1)}\!\right) &\!\!\!\!\!=\!\!\!\!\!&\!\!\sum_{n=1}^{\bar{N}_{\textrm{NDA}}}\!\!E_{a_m}\!\left\{\!L\big({\bm\theta}_k|a(n)=a_m\big)\bigg |\widehat{\bm{\theta}}_k^{(q-1)}\!\!,{\bf y}_k(n)\right\},\nonumber\\&\!\!\!\!\!\!\!\!\!\!&
\end{eqnarray}
where $E_{a_m}\{.\}$ is the expectation over all the possible transmitted symbols, $\{a_m\}_{m=1}^{M}$, and $\widehat{{\bm\theta}_k}^{(q-1)}$ is the estimated parameter vector at the $(q-1)^{th}$ iteration. After some algebraic manipulations, it can be shown that:
\begin{eqnarray}\label{Eq.50}
Q\left(\bm{\theta}_k|\widehat{{\bm \theta}_k}^{(q-1)}\right)&\!\!\!\!=\!\!\!\!&-\bar{N}_{\textrm{NDA}}N_r\ln(2\pi\sigma^2)-\frac{1}{2\sigma^2}\sum_{i=1}^{N_r}\!\Bigg(\!M_{2,k}^{(i)}+\nonumber\\
&\!\!\!\!\!\!\!\!& \!\!\!\!\!\!\!\!\sum_{n=1}^{\bar{N}_{\textrm{NDA}}}\!\alpha_{n,k}^{(q-1)}\big|{\bf c}_{i,k}^T{\bf t}(n)\big|^2\!\!-2\beta^{(q-1)}_{i,n,k}({\bf c}_{i,k})\!\Bigg),
\end{eqnarray}
where $M_{2,k}^{(i)}=E\{|y_{i,k}(n)|^2\}$ is the second-order moment of the received samples over the $i^{th}$ receiving antenna element and:
\begin{eqnarray}
\label{Eq.51a}\alpha_{n,k}^{(q-1)}&\!\!\!\!=\!\!\!\!&E_{a_m}\left\{|a_m|^2\bigg |\widehat{{\bm \theta}_k}^{(q-1)},{ \bf y}_k(n)\right\}\\
&\!\!\!\!=\!\!\!\!&\sum_{m=1}^MP_{m,n,k}^{(q-1)}|a_m|^2, \\
{\bf\beta}^{(q-1)}_{i,n,k}({\bf c}_{i,k})&\!\!\!\!=\!\!\!\!&E_{a_m}\!\!\left\{\!\Re\big\{y_{i,k}^*(n)a_m{\bf t}^T(n){\bf c}_{i,k}\big\}\bigg|\widehat{{\bm \theta}_k}^{(q-1)}\!\!,{ \bf y}_k(n)\right\}\nonumber\\
\label{Eq.51b}&\!\!\!\!=\!\!\!\!&\sum_{m=1}^MP_{m,n,k}^{(q-1)}\Re\left\{y_{i,k}^*(n)a_m{\bf t}^T(n){\bf c}_{i,k}\right\}.
\end{eqnarray}
In (\ref{Eq.51a}) and (\ref{Eq.51b}),  $P_{m,n,k}^{(q-1)}=P\left(a_m|{\bf y}_k(n);\widehat{{\bm \theta}_k}^{(q-1)}\right)$ is the {\it a posteriori} probability of $a_m$ at iteration $(q-1)$ which can be computed using the Bayes formula as follows:
\begin{eqnarray}\label{Eq.52}
P_{m,n,k}^{(q-1)}=\frac{P[a_m]P\left({\bf y}_k(n)\big|a_m;\widehat{{\bm \theta}_k}^{(q-1)}\right)}{P\left({\bf y}_k(n);\widehat{{\bm \theta}_k}^{(q-1)}\right)}.
\end{eqnarray}
Since the transmitted symbols are equally likely, we have $P[a_m]=1/M$, and thus:
\begin{equation}\label{Eq.53}
P\left({\bf y}_k(n);\widehat{{\bm \theta}_k}^{(q-1)}\right)=\frac{1}{M}\sum_{m=1}^MP\left({\bf y}_k(n)\big |a_m;\widehat{{\bm \theta}_k}^{(q-1)}\right).
\end{equation}
For normalized-energy constant-envelope constellations (such as MPSK), we have $|a_m|^2=1$ for all $a_m\in\mathcal{C}$ and, therefore, $\alpha_k^{(q-1)}(n)$ reduces simply to one (for all $n$) and does not need to be computed.
Now, the M-step can be fulfilled by determining the parameters that maximize the output of the E-step, obtained in (\ref{Eq.50}):
\begin{equation}\label{Eq.54}
\widehat{{\bm 
\theta}_k}^{(q)}=\arg \max_{\bm{\theta}_k}~Q\left({\bm{\theta}}_k\big |\widehat{\bm{\theta}_k}^{(q-1)}\right).
\end{equation}
At this stage, in order to avoid the cumbersome differentiation of the underlying objective function with respect to the complex vectors, $\{{\bf c}_{i,k}\}_{i=1}^{N_r}$, we split them into  ${\bf c}_{i,k}=\Re\{{\bf c}_{i,k}\}+j\Im\{{\bf c}_{i,k}\}$. We then  maximize instead $Q\left({\bm{\theta}}_k\big |\widehat{\bm{\theta}_k}^{(q-1)}\right)$ with respect to $\Re\{{\bf c}_{i,k}\}$ and $\Im\{{\bf c}_{i,k}\}$ yielding thereby, at the convergence of the iterative algorithm, their respective ML estimates $\Re\{\widehat{{\bf c}}_{i,k}\}$ and  $\Im\{\widehat{{\bf c}}_{i,k}\}$.  By the invariance principle of the ML estimator, we easily obtain the NDA ML estimate of ${\bf c}_{i,k}$ as $\widehat{\bf c}_{i,k}=\Re\{\widehat{{\bf c}}_{i,k}\}+j\Im\{\widehat{{\bf c}}_{i,k}\}$. Therefore,  using the fact that ${\bf t}(n)^T\big(\Re\{{\bf c}_{i,k}\}\Im\{{\bf c}_{i,k}\}^T -\Im\{{\bf c}_{i,k}\}\Re\{{\bf c}_{i,k}\}^T\big){\bf t}(n)=0~\forall~{\bf c}_{i,k}\in \mathbb{C}^L$ and after some algebraic manipulations, it can be shown that:
\begin{eqnarray}\label{Eq.71}
Q\!\left(\bm{\!\theta}_k|\widehat{{\bm \theta}_k}^{(q-1)}\right)&\!\!\!\!=\!\!\!\!&-\bar{N}_{\textrm{NDA}}N_r\ln(2\pi\sigma^2)\!
-\!\frac{1}{2\sigma^2}\!\sum_{i=1}^{N_r}\!\Bigg[\!M_{2,k}^{(i)}\!+\nonumber\\
&\!\!\!\!\!\!\!\!&\!\!\!\!\!\!\!\!\!\!\!\!\!\!\!\!\!\!\!\!\!\!\!\!\!\!\!\!\!\!\!\!\sum_{n=1}^{\bar{N}_{\textrm{NDA}}}\!\!\left(\!{\bf t}^T(n){\bf C}_{i,k}{\bf 
t}(n)\!-\!2\sum_{m=1}^M\!\!P_{m,n,k}^{(q-1)}{\widetilde{\bf c}^{(m)T}_{i,k}}
{\bf t}(n)\!\!\right)\!\!\Bigg].
\end{eqnarray}
where ${\bf C}_{i,k}$ and $\widetilde{\bf c}^{(m)}_{i,k}$ are, respectively, a  matrix and a  column vector that are explicitly constructed from the real and imaginary parts of ${\bf c}_{i,k}$ as follows:
\begin{eqnarray}
\!\!\!\!\!\!\!\!{\bf C}_{i,k}&\!\!\!\!=\!\!\!\!&\Re\{{\bf c}_{i,k}\}\Re\{{\bf c}_{i,k}\}^T +\Im\{{\bf c}_{i,k}\}\Im\{{\bf c}_{i,k}\}^T,\\
\!\!\!\!\!\!\!\!\widetilde{\bf c}^{(m)}_{i,k}&\!\!\!\!=\!\!\!\!&\Re\{y^*_{i,k}(n)a_m\}\Re\{{\bf c}_{i,k}\}+\Im\{y^*_{i,k}(n)a_m\}\Im\{{\bf c}_{i,k}\}.
\end{eqnarray}
After differentiating (\ref{Eq.71}) with respect to $\Re\{{\bf c}_{i,k}\}$ and $\Im\{{\bf c}_{i,k}\}$ and setting the resulting equations to zero, we obtain the NDA estimates of the real and imaginary parts of ${\bf c}_{i,k}$, at the $q^{th}$ iteration, as follows:
\begin{eqnarray}\label{Eq.72}
\Re\{\widehat{{\bf c}}^{(q)}_{i,k}\}\!\!\!\!\!\!&=&\!\!\!\!\!\!\left(\!\sum_{n=1}^{\bar{N}_{\textrm{NDA}}}\!\!{\bf t}(n){\bf t}^T\!(n)\!\!\right)^{\!\!\!\!-1}\!\!\!\!\left(\!\sum_{n=1}^{\bar{N}_{\textrm{NDA}}}\!\sum_{m=1}^M\!\!P_{m,n,k}^{(q-1)}\Re\{y_{i,k}^*(n)a_m\}{\bf t}(n)\!\!\right),\nonumber\\
\end{eqnarray}
and
\begin{eqnarray}\label{Eq.73}
\Im\{\widehat{{\bf c}}^{(q)}_{i,k}\}\!\!\!\!\!\!&=&\!\!\!\!\!\!\left(\!\sum_{n=1}^{\bar{N}_{\textrm{NDA}}}\!\!{\bf t}(n){\bf t}^T\!(n)\!\!\right)^{\!\!\!\!-1}\!\!\!\!\left(\!\sum_{n=1}^{\bar{N}_{\textrm{NDA}}}\!\sum_{m=1}^M\!\!P_{m,n,k}^{(q-1)}\Im\{y_{i,k}^*(n)a_m\}{\bf t}(n)\!\!\right).\nonumber\\
\end{eqnarray}
Then, using the identity $\widehat{{\bf c}}^{(q)}_{i,k}=\Re\{\widehat{{\bf c}}^{(q)}_{i,k}\}+j\Im\{\widehat{{\bf c}}^{(q)}_{i,k}\}$ and after some simplifications, we derive the expression of $\widehat{{\bf c}}^{(q)}_{i,k}$ as follows:
\begin{eqnarray}\label{Eq.55}
\widehat{{\bf c}}^{(q)}_{i,k}=\left(\sum_{n=1}^{\bar{N}_{\textrm{NDA}}}{\bf t}(n){\bf t}^T(n)\right)^{-1}\left(\sum_{n=1}^{\bar{N}_{\textrm{NDA}}} \lambda_{i,n,k}^{(q-1)}{\bf t}(n)\right),
\end{eqnarray}
in which $\lambda_{i,n,k}^{(q-1)}$ is given by:
\begin{eqnarray}\label{lambda}
\lambda_{i,n,k}^{(q-1)}=[\widehat{a}^{(q-1)}_k(n)]^*y_{i,k}(n)
\end{eqnarray}
where
\begin{eqnarray}\label{Eq.56_1}
\widehat{a}^{(q-1)}_k(n)&=&\sum_{m=1}^MP_{m,n,k}^{(q-1)}a_m,
\end{eqnarray}
is the previous \textit{soft} estimate for the unknown transmitted symbol, $a_k(n)$, involved in (\ref{Eq.45}). Lastly, by  differentiating  (\ref{Eq.71}) with respect to $\sigma^2$, setting the resulting equation to zero, and replacing therein ${\bf c}_{i,k}$ by $\widehat{{\bf c}}^{(q-1)}_{i,k}$, we obtain a new estimate of the noise power at the $q^{th}$ iteration as follows: 
\begin{eqnarray}\label{Eq.57}
2\widehat{\sigma^2}^{(q)}_k=\frac{\sum_{i=1}^{N_r}\left(M_{2,k}^{(i)}+\eta_{i,k}^{(q-1)}\right)}{\bar{N}_{\textrm{NDA}}N_r},
\end{eqnarray}
where:
\begin{eqnarray}\label{Eq.58}
\eta_{i,k}^{(q-1)}&\!\!\!\!=\!\!\!\!&\sum_{n=1}^{\bar{N}_{\textrm{NDA}}}\bigg[{\bf t}^T(n)\left({\bf \widehat{c}}^{(q-1)}_{i,k}\right)^*\left({\bf \widehat{c}}^{(q-1)}_{i,k}\right)^T{\bf t}(n)+\nonumber\\
&\!\!\!\!\!\!\!\!&~~~~~~~~~~~~~~~~~~~~~~~~~~ \alpha_{n,k}^{(q-1)}-2\beta^{(q-1)}_{i,n,k}\left(\widehat{{\bf c}}^{(q-1)}_{i,k}\right)\bigg]\nonumber\\
&\!\!\!\!=\!\!\!\!&\sum_{n=1}^{\bar{N}_{\textrm{NDA}}}\bigg[\big|{\bf t}^T(n)\widehat{{\bf c}}_{i,k}^{(q-1)}\big|^2+\alpha_{n,k}^{(q-1)}-2\beta^{(q-1)}_{i,n,k}\left(\widehat{{\bf c}}^{(q-1)}_{i,k}\right)\bigg].\nonumber\\
&\!\!\!\!\!\!\!\!&
\end{eqnarray}
After few iterations (i.e., in the range of 10) and with careful initialization, the EM algorithm converges  over each $k^{th}$ approximation window  to the exact NDA ML estimates $\widehat{{\bf c}}^{(k)}_{i,\textrm{NDA}}$ and $\widehat{\sigma^2_k}_{\textrm{,NDA}}$. The latter is then averaged over all the local approximation windows to obtain a more refined estimate as follows:
\begin{eqnarray}\label{averaged_noise_estimate}
\widehat{\sigma^2}_{\textrm{NDA}}=\frac{\bar{N}_{\textrm{NDA}}}{N}\sum_{k=1}^{N/\bar{N}_{\textrm{NDA}}}\widehat{\sigma_k^2}_{,\textrm{NDA}}.
\end{eqnarray}
Finally, given (\ref{Eq.55}) and (\ref{averaged_noise_estimate}), and taking into account all the approximation windows of size $\bar{N}_{\textrm{NDA}}$ within the same observation window of size $N$, the NDA ML SNR estimator is obtained as:
\begin{equation}\label{Eq.59}
\widehat{\rho}_{i,\textrm{NDA}}=\frac{\sum_{k=1}^{N/\bar{N}_{\textrm{NDA}}}\sum_{n=1}^{\bar{N}_{\textrm{NDA}}}|\widehat{a}_k(n)|^2\big|{\bf t}^T(n)\widehat{{\bf c}}^{(k)}_{i,\textrm{NDA}}\big |^2}{N\left(2\widehat{\sigma^2}_{\textrm{NDA}}\right)},
\end{equation} 
where $\widehat{a}_k(n)$ is the final (i.e., at the convergence) \textit{soft} estimate of the $n^{th}$ transmitted symbol, $a_k(n)$, within the $k^{th}$ approximation window.
\subsection{Appropriate initialization of the iterative EM algorithm using the DA estimator}
 Recall that the EM algorithm is iterative in nature and, therefore, its performance is closely tied to the initial guess $\widehat{{\bm\theta}_k}^{(0)}$ within each approximation window. We will see in the next section that when it is not appropriately initialized, its performance is indeed severely affected, especially at high SNR levels. This is actually a serious problem inherent to any iterative algorithm whose objective function is not convex (i.e., multimodal). That is, it may settle on any local maximum if it happens that the algorithm is accidentally initialized close to it. Fortunately, an appropriate initial guess about the polynomial coefficients, ${\bf \widehat{c}}_{i,k}^{(0)}$, and the noise variance, $\widehat{\sigma^2}^{(0)}$, can be locally acquired using very few pilot symbols by applying the DA ML estimator  developed in the previous section.\\
In order to initialize the EM algorithm with the DA estimates, we proceed as follows. Using the pilot symbols only, we begin by estimating the local polynomial coefficients, $\{{\bf \widehat{c}}_{i,{\textrm{DA}}}^{(k)}\}_{k}$, using the DA estimator over  approximation windows of size $\bar{N}_{\textrm{DA}}$ (possibly different from $\bar{N}_{\textrm{NDA}}$).  In Section III, ${\bf \widehat{c}}_{i,{\textrm{DA}}}^{(k)}$ was multiplied by the matrix ${\bf T}'$ in order to obtain, over each $k^{th}$ approximation window, the DA estimates for the channel coefficients, ${\bf \widehat{h}}'^{(k)}_{i,\textrm{DA}}$, at pilot positions only  \big(i.e., ${\bf \widehat{h}}'^{(k)}_{i,\textrm{DA}}={\bf T}'{\bf \widehat{c}}_{i,{\textrm{DA}}}^{(k)}$\big). Yet, they can also be multiplied by another matrix ${\bf T}_{\bar{N}_{\textrm{DA}}}$ in order to obtain the pilot-based estimates for the channel coefficients at both pilot and non-pilot positions over  each DA approximation window \big(i.e., ${\bf \widehat{h}}^{(k)}_{i,\textrm{DA}}={\bf T}_{\bar{N}_{\textrm{DA}}}{\bf \widehat{c}}_{i,{\textrm{DA}}}^{(k)}$\big). The underlying  time matrix ${\bf T}_{\bar{N}_\textrm{DA}}$ is equivalent to ${\bf T}_{\bar{N}_{\textrm{NDA}}}$ in (\ref{Eq.39}) except the fact that it contains $\bar{N}_{\textrm{DA}}$ instead of $\bar{N}_{\textrm{NDA}}$ rows. Then, over each $i^{th}$ antenna element, the obtained pilot-based estimates, $\left\{{\bf \widehat{h}}^{(k)}_{i,\textrm{DA}}\right\}_k$, are stacked together to form a single vector,  $\widehat{\bf h}_{i,\textrm{DA}}$, that contains all the pilot-based estimates of the channel coefficients over the entire observation window. The latter is again divided into several adjacent and disjoint blocks, $\widehat{\bf h}^{(k)}_{i,\textrm{DA}}$, each of which  is now of size $\bar{N}_{\textrm{NDA}}$ \big(instead of $\bar{N}_{\textrm{DA}}$ in the DA scenario\big). Then, according to (\ref{Eq.44}), the initial guess about the polynomial coefficients --- within each $k^{th}$ local NDA approximation window --- is obtained from the $k^{th}$  block using:
\begin{eqnarray}
\widehat{{\bf c}}^{(0)}_{i,k}&=&\left({\bf T}_{\bar{N}_{\textrm{NDA}}}^T{\bf T}_{\bar{N}_{\textrm{NDA}}}\right)^{-1}{\bf T}_{\bar{N}_{\textrm{NDA}}}^T~\widehat{\bf h}^{(k)}_{i,\textrm{DA}}\nonumber\\
&&~~~~~~~~~~~~~~~\textrm{for} ~~~k=1,2,\cdots, N/\bar{N}_{\textrm{NDA}}.
\end{eqnarray}
The initial guess about the noise variance is simply $\widehat{\sigma^2}^{(0)}=\widehat{\sigma^2}_{\textrm{DA}}$ obtained in (\ref{noise_variance_DA}). In the following,  we will use two different designations for the new EM-based estimator depending on the initialization procedure. We shall refer to it as ``\textit{completely-NDA}'' if  initialized \textit{arbitrarily} and as ``\textit{hybrid}'' when  initialized \textit{appropriately} using the DA estimator.   We will also use two different designations for the DA estimator.  We shall refer to it as ``\textit{pilot-only DA}'' when applied using the pilot symbols only (which are $N/N_p$ out of the $N$ transmitted symbols with $N_p>1$); and as ``\textit{completely DA}'' when applied in  another scenario in which all the $N$ transmitted symbols are assumed to be  perfectly known, i.e., $N_p=1$. This  scenario is encountered in  many modern communication systems which have a small CRC (at the PHY layer) serving as a stopping criterion for turbo code detection. This means that at the end of the decoding process, the system can recognize whether the bits were detected correctly or not (i.e., if the CRC matches or not). Thus, at the output of the decoder, one has access to the transmitted information bits from which all the transmitted channel symbols can be easily obtained. These decoded symbols are then used as pilots for the DA estimator in a ``\textit{completely DA}'' mode. Moreover, in some radio interface technologies such as CDMA, a code-multiplexed  pilot channel is considered with a completely known data sequence. In OFDM transmissions, as well, some carriers might bear completely known data sequences.  
\subsection{EM-based ML SNR estimation with hard symbol detection}
The EM-based SNR estimator developed in section IV-A relies on the \textit{soft detection} (SD) of the transmitted symbols as seen from (\ref{Eq.56_1}). In fact, at each time instant $n$,  all the constellation points are scanned and the corresponding \textit{a posteriori} probabilities (APPs), $P_{m,n,k}$, are updated from one iteration to another. With a properly selected setup\footnote{This amounts to carefully choosing  the local approximation window sizes ($\bar{N}_{\textrm{NDA}}$ and $\bar{N}_{\textrm{DA}}$) pertaining, respectively, to  the ``\textit{hybrid}'' SNR estimator and the DA version used to initialize it; choices that both depend on the normalized Doppler frequency $F_DT_s$ as established and reported in table I at the end of the next section.}, the \textit{hybrid} EM-based estimator always converges  to the global maximum of the LLF for moderate-to-high SNR values. Therefore, over that SNR region and at the convergence of the algorithm, the APPs of the wrong symbols are \textit{almost} equal to zero. As such, the weighted sum involved in (\ref{Eq.56_1}) returns a very accurate \textit{soft} estimate, $\widehat{a}_k(n)$, of  the   actual $n^{th}$ transmitted symbol  (over each $k^{th}$ local approximation window). This makes the ``\textit{hybrid}'' EM-based SNR estimator equivalent in performance to the ``\textit{completely DA}'' \textit{biased}  estimator. Therefore, the same bias-correction procedure highlighted earlier in (\ref{Eq.26}) can be exploited here  using $\epsilon=L/\bar{N}_{\textrm{NDA}}$.  To be more specific, we will further refer to the ``\textit{completely-NDA}'' and ``\textit{hybrid}'' EM-based estimators as ``\textit{completely-NDA-SD}'' and ``\textit{hybrid-SD}'' when they are applied with \textit{soft} detection (SD) using (\ref{Eq.56_1}).\\
Yet, for low SNR values,  \textit{soft} detection  may not be optimal and hence both the ``\textit{completely-NDA-SD}'' and ``\textit{hybrid-SD}'' EM-based estimators are expected to  depart from the ``\textit{completely DA}'' estimator. Therefore, one may resort to  \textit{hard} detection (HD)  in order to bridge such performance gap. In a nutshell, HD is a separate task that may be applied iteratively (i.e., at each $q^{th}$ iteration) by  taking each of the soft estimates, $\widehat{a}_k^{(q)}(n)$, in (\ref{Eq.56_1}) as input to return its closest symbol, $\bar{a}^{(q)}_k(n)$, in the constellation alphabet:
\begin{eqnarray}\label{hard_detection}
\bar{a}^{(q)}_k(n)=\underset{a_m \in {\mathcal{C}}}\argmin\left|a_m-\widehat{a}^{(q-1)}_k(n)\right|^2. 
\end{eqnarray}
Then, $\bar{a}^{(q)}_k(n)$ is used in (\ref{lambda}) instead of $\widehat{a}_k^{(q)}(n)$. When applied with iterative hard detection (IHD),  the ``\textit{completely-NDA}'' and ``\textit{hybrid}'' EM-based estimators are referred to as ``\textit{completely-NDA-IHD}'' and ``\textit{hybrid-IHD}'', respectively. One  other option would be to apply the HD task  only once at the convergence of the algorithm [i.e., final \textit{hard} detection (FHD)]. In this case, (\ref{hard_detection}) is applied on the  \textit{soft} symbols' estimates obtained at the very last iteration only. Hence, we drop the iteration index $q$ in the output,  $\bar{a}_k(n)$, of (\ref{hard_detection}) which is reinjected instead of $\widehat{a}_k(n)$ obtained at the convergence. When applied with FHD, the two versions of the EM-based estimator are designated, respectively, as ``\textit{completely-NDA-FHD}'' and ``\textit{hybrid-FHD}''.   
Finally, the multiple capabilities of the proposed NDA ML SNR estimator to implicitly and simultaneously i) identify the time-varying channel coefficients, ii) estimate the noise power, and iii) detect or demodulate the transmitted symbols  owe to be underlined. Yet study and assessment of these capabilities or functionalities (i.e., channel identifier, noise power estimator, and  data demodulator or detector) fall beyond the scope of this paper.
\section{Simulation Results}
\begin{figure}[!t]
\vskip -0.25cm
\begin{centering}
\includegraphics[scale=0.57]{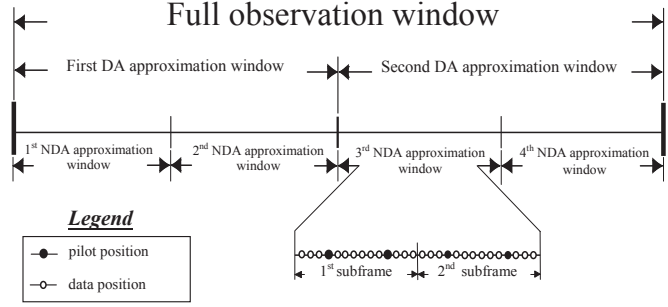}
\caption{Data and pilot symbols layout with  four and two DA and NDA local approximation windows, respectively, with $N=112$, $\bar{N}_{\textrm{DA}}=56$, and $\bar{N}_{\textrm{NDA}}=28$.} 
\label{explanatory}
\end{centering}
\end{figure}

\begin{figure}[!t]
\begin{centering}
\includegraphics[scale=0.52]{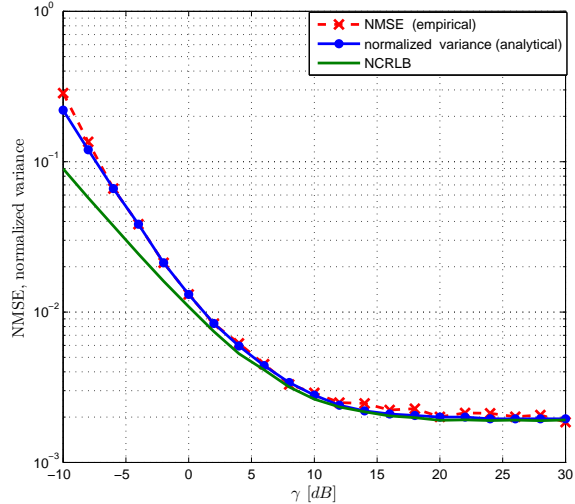}
\caption{ NMSE (empirical) and normalized variance (analytical) of the \textit{unbiased} DA ML  estimator vs. the average SNR $\gamma$, with $N_r=2$, $N=112$, $\bar{N}_{\textrm{DA}}=112$, $\bar{N}_{\textrm{NDA}}=N/2=56$, $F_DT_s=7\times 10^{-3}$ and $L=4$, 16-QAM.} 
\label{variance_and_NMSE_DA}
\end{centering}
\end{figure}

\begin{figure}[!t]
\begin{centering}
\includegraphics[scale=0.45]{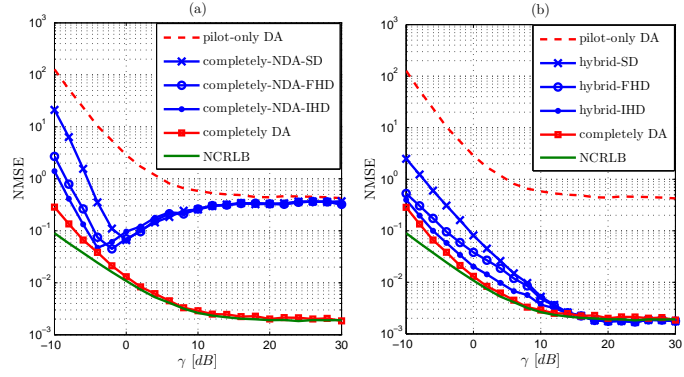}
\caption{ NMSE of (a): ``\textit{completely-NDA}'' and (b): ``\textit{hybrid}'' EM-based estimators against benchmarks vs. the average SNR $\gamma$, with $N_r=2$, $N=112$, $\bar{N}_{\textrm{DA}}=112$, $\bar{N}_{\textrm{NDA}}=N/2=56$, $F_DT_s=7\times 10^{-3}$ and $L=4$, 16-QAM.} 
\label{DA_pilots}
\end{centering}
\end{figure}

\begin{figure}[!t]
\begin{centering}
\includegraphics[scale=0.45]{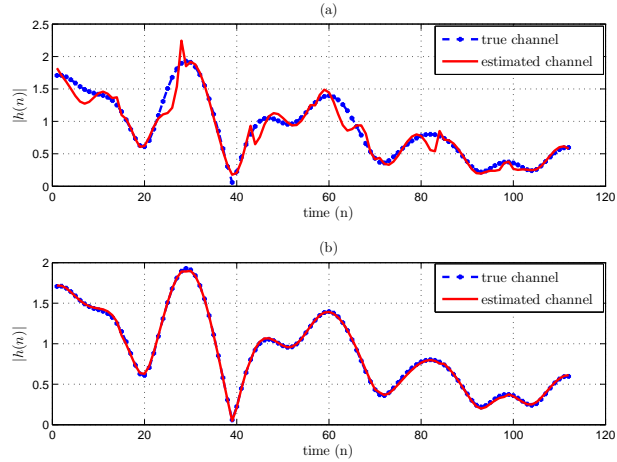}
\caption{True vs. estimated channel magnitude for the EM-based algorithm when initialized (a) arbitrarily with ones, and (b) appropriately with the ``\textit{pilot-only DA}'' estimates, for $F_DT_s=3.5\times 10^{-2}$, $N=112$, $\bar{N}_{\textrm{DA}}=28$, $\bar{N}_{\textrm{NDA}}=14$, and $L=4$.} 
\label{initialization}
\end{centering}
\end{figure}

\begin{figure}[!t]
\begin{centering}
\includegraphics[scale=0.45]{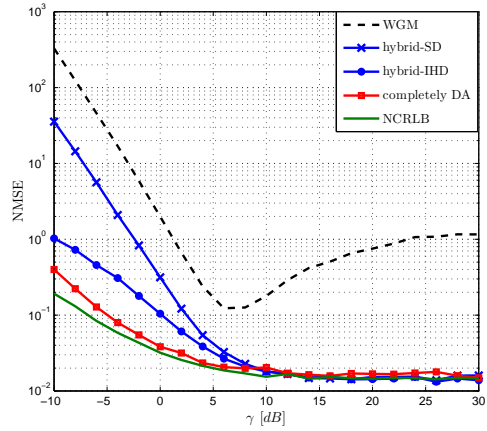}
\caption{Comparison of our new SNR estimators with WGM over SISO systems, i.e., $N_r=1$ with $F_DT_s=7\times 10^{-3}$, QPSK.} 
\label{SISO}
\end{centering}
\end{figure}

\begin{figure}[!t]
\begin{centering}
\includegraphics[scale=0.5]{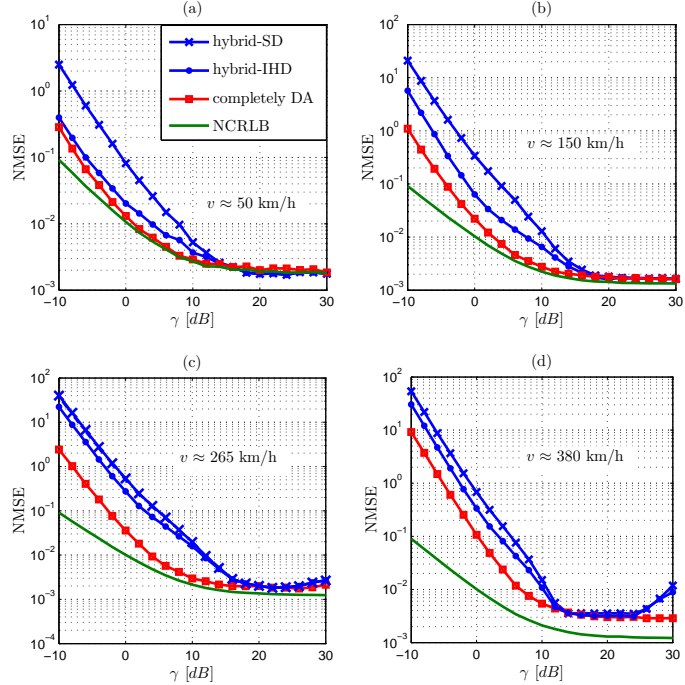}
\caption{NMSE for the ``\textit{hybrid}''  EM-based and the ``\textit{completely DA}'' 
unbiased estimators vs. the average SNR  with $N=112$ and $N_r=2$ for: (a) $F_DT_s=7\times 10^{-3}$, $\bar{N}_{\textrm{DA}}=112$, $\bar{N}_{\textrm{NDA}}=56$, (b) $F_DT_s=2\times 10^{-2}$, $\bar{N}_{\textrm{DA}}=28$, $\bar{N}_{\textrm{NDA}}=28$, (c) $F_DT_s=3.5\times 10^{-2}$, $\bar{N}_{\textrm{DA}}=28$, $\bar{N}_{\textrm{NDA}}=14$ and (d) $F_DT_s=5\times 10^{-2}$, $\bar{N}_{\textrm{DA}}=14$, $\bar{N}_{\textrm{NDA}}=7$, 16-QAM.} 
\label{dopplers}
\end{centering}
\end{figure}

In this section, we assess the performance of our new DA and NDA ML \textit{instantaneous} SNR estimators. All the presented results are obtained by running extensive Monte-Carlo simulations over $5000$ realizations. The estimators' performance is evaluated in terms of the normalized (by the average SNR) mean square error (NMSE) and compared to the normalized CRLB (NCRLB) defined as:
\begin{equation}\label{Eq.60}
\textrm{NMSE}(\widehat{\rho_i})=\frac{E\{(\rho_i-\widehat{\rho_i})^2\}}{\gamma^2},~\textrm{NCRLB($\rho_i$)}=\frac{\textrm{CRLB($\rho_i$)}}{\gamma^2},\nonumber
\end{equation}
where $\gamma=E\{|a(n)|^2\}/(2\sigma^2)$ is the average SNR per symbol.  Since the constellation energy is assumed to be normalized to one, i.e., $E\{|a(n)|^2\}=1$, $\gamma$ is simply given by $\gamma=1/(2\sigma^2)$. For the sake of complying with a practical and timely scenario, all the simulations  are conducted in the specific context of  uplink  LTE \cite{ref17}.  According to its signalling standard  specifications, two pilot OFDM symbols are inserted at the fourth and eleventh positions within the time-frequency grid of each subframe (consisting of $14$ OFDM symbols). In this way a pilot symbol is transmitted every seven OFDM symbols corresponding to $N_p=7$.
In Fig. \ref{explanatory}, we illustrate the data/pilot symbols layout adopted over each carrier considering an observation window of eight consecutive subframes (i.e., $N=112$), with typical choices of the DA and NDA local approximation window sizes $\bar{N}_{\textrm{DA}}=56$ and $\bar{N}_{\textrm{NDA}}=28$.
\\ In the sequel, the ``\textit{instantaneous}'' SNR estimation results are presented for the first subcarrier only, but they actually hold the same irrespectively of the subcarrier index. Moreover, all the results are obtained for \textit{complex} channels since, in practice, the baseband-equivalent representation of the channel coefficients in the discrete model (\ref{Eq.1}) is always complex. We will also consider QPSK and 16-QAM as representative examples for  constant-envelope and non-constant-envelope constellations, respectively.
First, we verify from Fig. \ref{variance_and_NMSE_DA} that the analytical variance of the \textit{unbiased} ML DA estimator which we developed in (\ref{Eq.28}) coincides with its NMSE computed empirically using Monte-Carlo computer simulations. The small discrepancies between them is due to lack of averaging.
 \\In Fig. \ref{DA_pilots}, we plot the NMSE for the ``\textit{completely-NDA}'' and ``\textit{hybrid}'' EM-based estimators (both with SD, IHD and FHD) and compare them to the ``\textit{pilot-only DA}'' and ``\textit{completely DA}'' estimators.
\\First, by closely inspecting Fig. \ref{DA_pilots}(a),  as expected intuitively  due to the fast time variations of the channel,  the ``\textit{pilot-only DA}'' estimator is not able to accurately estimate the SNR by relying  solely on the pilot symbols.  Therefore, the received samples at non-pilot positions must be exploited as well in order to account for the channel variations between the pilot positions. The ``\textit{completely-NDA}'' EM-based estimator does so and as such is able to provide substantial performance gains at low-to-medium SNR values against the ``\textit{pilot-only DA}'' method. Yet, its performance deteriorates severely at high SNR levels due to its initialization issues. This is where  the ``\textit{pilot-only DA}'' estimator actually becomes extremely useful even though its overall performance is not satisfactory. Indeed, its estimates are accurate enough to serve  as initial guesses for the ``\textit{hybrid}'' EM-based algorithm to make it converge to the global maximum of the LLF reaching thereby the CRLB as seen from Fig. \ref{DA_pilots}(b). To clearly show the effect of both \textit{arbitrary} and \textit{appropriate} initializations on the EM-based algorithm \big(i.e., the ``\textit{completely-NDA}'' and ``\textit{hybrid}'' estimators, respectively\big), we plot  in Fig. \ref{initialization} the corresponding true and estimated channel coefficients at an average SNR $\gamma=20$ dB.
\\Clearly, when initialized with the  ``\textit{pilot-only DA}'' estimates\footnote{See  section IV-B for more details about the pilot-assisted initialization process.}, the  iterative algorithm is able to  track the channel variations more accurately. Therefore, as clearly seen from Fig. \ref{DA_pilots}(b), the ``\textit{hybrid}'' EM-based SNR estimator exhibits paramount performance improvements  especially for moderate to high SNR levels. Fig. \ref{DA_pilots}(b) also highlights the advantage of performing  IHD since the ``\textit{hybrid-IHD}'' EM-based estimator is almost equivalent, over the entire SNR range, to the ``\textit{completely DA}''  estimator which assumes all the symbols to be perfectly known. Even more, both estimators ultimately coincide  with the CRLB  which quantifies theoretically the best achievable performance ever.  Fig. \ref{DA_pilots}(b) also reveals that IHD yields more accurate SNR estimates than FHD and, therefore, the latter will not be considered in the remaining simulations.  The ``\textit{completely-NDA}'' EM-based estimator with SD, IHD, and FHD  was also included in Fig. \ref{DA_pilots}(a) to have these preliminary comparisons exhaustive and  to motivate  the use of the ``\textit{pilot-only}'' DA estimates in initialization. Thus, in the remaining simulations we will focus on the ``\textit{hybrid}'' EM-based estimator with SD and IHD only. Yet, we will keep using the ``\textit{completely DA}''  estimator and the CRLB as ideal benchmarks.\\     
Now, we will compare our new ``\textit{hybrid}'' estimator against the only reported work\footnote{Note also that, using exhaustive computer simulations, we have demonstrated the clear superiority of our new  ML estimators against other state-of-the-art techniques developed for constant channels [\ref{ref19}, \ref{ref8}, \ref{ref22}] and time-varying channels [\ref{Morelli}, \ref{Faouzi}]. The results were not included in this paper due to lack of space.} on  EM-based ML SNR estimation over time-varying channels introduced by A. Wiesel \textit{et al.} in  [\ref{ref14}]. Using the initials of its authors' names, we will henceforth designate it as ``WGM''. This estimator was originally derived for  single-input single output (SISO) systems. Thus, it can be directly applied at the output of each antenna element in order to estimate the \textit{instantaneous} SNR in SIMO configurations. Yet, it can also be easily modified to take advantage of the antenna gain offered by  SIMO systems experiencing \textit{uniform} noise. In fact, over each $i^{th}$ antenna branch, the SISO WGM algorithm yields two estimates; one for the signal power, $\widehat{P}_i$, and the other for the noise power, $\widehat{N}_0^{(i)}$. The individual estimates  $\{\widehat{N}_0^{(i)}\}_{i=1}^{N_r}$ can be averaged over the $N_r$ receiving antenna elements to provide a more refined estimate, $\widehat{N}_0$, for the unknown noise power. The SIMO-enhanced WGM estimator over each antenna branch, referred to hereafter as the ``WGM-SIMO'' estimator, is then redefined as $\widehat{\rho}_i=\widehat{P}_i/\widehat{N}_0$.
\begin{figure}[!t]
\begin{centering}
\includegraphics[scale=0.5]{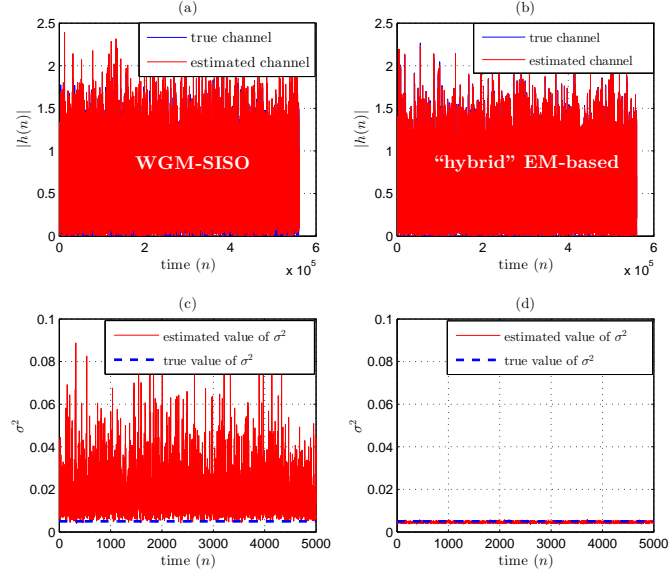}
\caption{True and estimated channel amplitude and noise variance for: WGM estimator (left-hand side) and our ``\textit{hybrid}'' EM-based estimator  with $N_r = 1$, i.e., SISO (right-hand side) at an average SNR $\gamma=20$ dB (i.e., $\sigma^2=0.005$) and $F_DT_s=7\times 10^{-3}$, QPSK.} 
\label{comparaison_of_channel_and_noise_variance_estimates}
\end{centering}
\end{figure}
\\In Fig. \ref{comparaison_of_channel_and_noise_variance_estimates}, we compare  our ``\textit{hybrid}'' EM-based estimator (with $N_r=1$, i.e., SISO) against WGM in terms of \textit{complex} channel tracking capabilities and noise variance estimation accuracy over 5000 Monte-Carlo runs (i.e., 5000 consecutive observation windows each of size $N=112$).
The reason behind considering such a very large number of observation windows --- although it does not allow one to distinguish the true channel from its estimates --- is to show that our estimator always converges to the global maximum. This can be, in fact, easily deduced by inspecting the noise variance estimates in the same figure. In plain English, under \textit{complex} time-varying channels, the multidimensional LLF has  many local maxima (i.e., multimodal) and the WGM estimator gets trapped into one of them due to its initialization issues. Therefore, as seen  from Fig. \ref{comparaison_of_channel_and_noise_variance_estimates}(c), it  is not able to estimate the noise variance over \textit{almost all} the observation windows. Owing to our new \textit{proper initialization procedure}, however, our ``\textit{hybrid}'' EM-based estimator enjoys guaranteed global optimality and thus returns very accurate noise variance estimates over \textit{all} the observation windows. Consequently, in contrast to WGM,  it  achieves the DA CRLB as shown in Fig. \ref{SISO}. Most remarkably, the ``\textit{hybrid}'' algorithm is able to do so with $86~\%$ of the transmitted symbols being completely unknown (corresponding to a pilot insertion rate of $1/N_p=1/7$ as advocated by the signalling standard specifications  of the LTE uplink).
\begin{figure}[!t]
\begin{centering}
\includegraphics[scale=0.4]{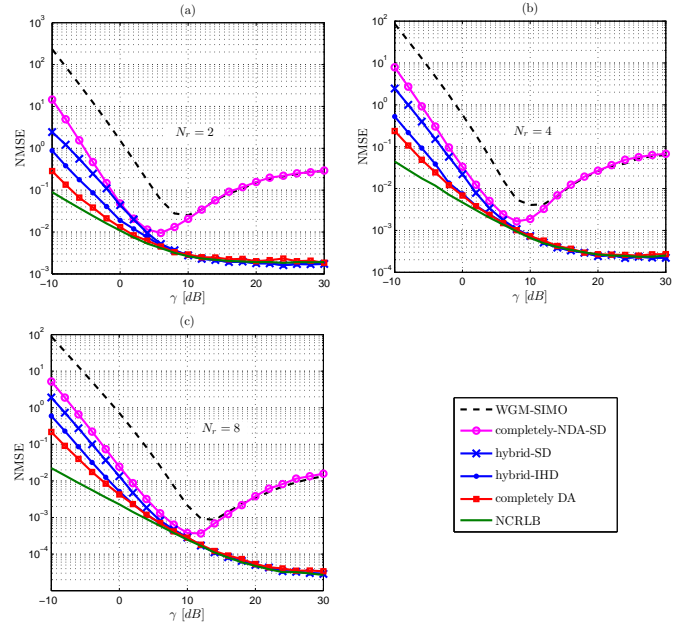}
\caption{Comparison of our estimators against WGM-SIMO  for different numbers of receiving antenna elements: (a) $N_r=2$, (b) $N_r=4$, and (c) $N_r=8$, with $F_DT_s=7\times 10^{-3}$, $N=112$,  $\bar{N}_{\textrm{DA}}=112$, and $\bar{N}_{\textrm{NDA}}=56$, $L=4$,  QPSK.} 
\label{wiesel_simo}
\end{centering}
\end{figure}
\\Fig. \ref{wiesel_simo} depicts  the performance of WGM-SIMO and the different versions of our estimator over three SIMO configurations (i.e., $N_r=2,~4,~\textrm{and}~8$). First, by inspecting  the behaviour of the WGM estimator across the three subfigures, it is  seen that the performance of its SIMO-enhanced version improves remarkably with the number of receiving antenna elements. For instance, at the typical value of the average SNR $\gamma=30$ dB,  it is seen from Figs. \ref{wiesel_simo}(a) and (b) that the variance of this estimator is reduced by a factor of $1/5$ when the number of antennae is doubled from $N_r=2$ to $N_r=4$.  The same improvements  hold --- although with a slightly smaller factor of $1/4$ --- by further doubling the array size from $N_r=4$ to $N_r=8$. Such improvements are actually due to the antennae \textit{gain} only. Indeed, since WGM-SIMO is not able to exploit the antenna \textit{diversity}, it is substantially outperformed even by our ``completely-NDA-SD'' estimator, for low-to-medium SNR levels. Here, we make a clear difference between the two concepts of antennae \textit{gain} and \textit{diversity}. The former is actually inherent to all SIMO systems experiencing uniform noise across the antenna elements \big(under correlated or uncorrelated channels\big). In this case, averaging the $N_r$ independent estimates of the same noise power produces a new estimate whose variance is always shrunk by a factor of $1/N_r$, improving thereby the final estimates of the per-antenna SNRs.\\
 Antennae diversity, however, is another more interesting feature of SIMO systems. Fully exploiting the antennae diversity consists in optimally combining the multiple independently-fading copies of the received signal in order to detect each of the transmitted symbols correctly.  By solving the ML criterion, our ``\textit{hybrid-SD}'' \big(or ``\textit{hybrid-IHD}''\big) EM-based estimator takes indeed advantage of the available spatial diversity to accurately estimate (or detect)  the unknown transmitted symbols. For these reasons and owing to our proper initialization procedure, the ``\textit{hybrid}'' EM-based algorithm (with SD or IHD)  outperforms by far WGM-SIMO over the entire SNR range.
From another perspective, the performance improvements  that are obtained by fully  exploiting the antennae \textit{gain} together with the antennae \textit{diversity} offered by SIMO  over SISO systems can be easily appreciated by comparing Figs. \ref{wiesel_simo} and \ref{SISO}. For instance, at the typical average SNR value of $\gamma=30$ dB, the NMSE of the ``\textit{hybrid}'' EM-based estimator is substantially reduced by a factor as high as $2500$  using  $8$ antenna branches compared to SISO. 
\\ So far, all the simulations where conducted under a normalized Doppler frequency of $F_DT_s=7\times 10^{-3}$ corresponding to a maximum Doppler shift $F_D\approx 100$ Hz with the sampling rate of LTE systems $T_s=71.42~\mu$s. This translates into  a medium user velocity $v=\frac{F_D}{F_c}c\approx 50$ Km/h at a carrier frequency  $F_c=2$ GHz with $c=3\times 10^8~m/s$ being the speed of light. Therefore,  we plot in Fig. \ref{dopplers} the performance of the newly derived ML estimator for higher normalized Doppler frequencies.  
\\It is seen from this figure that both the ``\textit{completely DA}'' and ``\textit{hybrid}'' estimators succeed in accurately estimating the SNR reaching thereby the DA CRLB  even at high Doppler frequencies. In Fig. \ref{dopplers}(d), for instance, the normalized Doppler frequency is as high as $F_DT_s=5\times 10^{-2}$ corresponding to a maximum Doppler frequency of $700$ Hz (translating to a user velocity as high as $v=380$ Km/h at $F_c=2$ GHz). Within the same context, we emphasize the fact that the sizes of the local approximation windows, $\bar{N}_{\textrm{NDA}}$ and $\bar{N}_{\textrm{DA}}$, for both the ``\textit{hybrid}'' estimator and the ``\textit{pilot-only DA}'' that is used to initialize it should be properly  selected according to the Doppler range as shown in  Table I.  
\begin{table}[!h]
\centering
\label{tableI}
\caption{local estimation configurations for different ranges of $F_DT_s$.}
\begin{tabular}{c|c|c|c|r|}
 \cline{2-5}
  & $\bar{N}_{\textrm{DA}}$ & $\bar{N}_{\textrm{NDA}}$ & $L_{\textrm{DA}}$ & $L_{\textrm{NDA}}$\\
  \hline
  \multicolumn{1}{ |c| }{$F_DT_s\leq 7\times 10^{-3}$} & 112 & 56 & 4 & 4 \\
  \hline
  \multicolumn{1}{ |c| }{$7\times 10^{-3}\leq F_DT_s\leq 2\times 10^{-2}$} & 28 & 28 & 4 & 4 \\
  \hline
  \multicolumn{1}{ |c| }{$2\times 10^{-2}\leq F_DT_s\leq 3.5\times 10^{-2}$} & 28 & 14 & 4 & 4 \\
  \hline
  \multicolumn{1}{ |c| }{$F_DT_s\geq 5\times 10^{-2}$} & 14 & 7 & 2 & 4 \\
  \hline
\end{tabular}
\end{table}
In practice, the Doppler frequency can be estimated from the samples received at the pilot positions and  the approximation window sizes are then selected accordingly.
When designing  these Doppler-dependent configurations, our primary goal was to obtain the lowest possible polynomial orders $L_{\textrm{DA}}$ and $L_{\textrm{NDA}}$ which define the sizes of the two matrices that need to be inverted. Yet, it should be mentioned that these small-size matrices are predefined ones. Hence, in practice, they can be computed and inverted offline  once for all, stored in memory, and then used in online estimation at no extra computational cost.
\section{Conclusion}
In this paper, we formulated and derived  ML estimators for the \textit{instantaneous} SNR  over time-varying SIMO channels using  local polynomial-in-time expansions. In the DA scenario, the ML estimator was derived in  closed form, and so were its bias, its variance and  the DA CRLB. In the NDA case, however, we proposed a ML solution that is based on the iterative EM concept  and that is able to converge to the global maximum within very few iterations. Appropriate initialization is indeed guaranteed by applying the DA estimator over periodically inserted pilot symbols. Furthermore, the new estimator is  applicable to any channel fading type over a relatively large Doppler range
 and for any linearly-modulated signal (i.e., PSK, PAM, QAM). Finally, it is able to reach the CRLB over a wide SNR range and  outperforms  by far  the new SIMO-extended version of the only  work published so far, to the best of our knowledge,  on EM-based ML SNR estimation over SISO time-varying channels.
\section*{Appendix A\\Proof of Theorem 1}
To begin with we define, $\widetilde{\mathbf{P}}_k$, as
the orthogonal projector on the signal subspace (of each $i^{th}$  antenna element) corresponding to the $k^{th}$ local DA approximation window (of size $\bar{N}_{\textrm{DA}}$) as follows:
\begin{eqnarray}
\widetilde{\mathbf{P}}_k=\bm{\Phi}_k(\bm{\Phi}_k^H\bm{\Phi}_k)^{-1}\bm{\Phi}_k^{H},
\end{eqnarray}
where  $\bm{\Phi}_k\triangleq\mathbf{A}_k\mathbf{T}$ and $\mathbf{A}_k$ is  a diagonal matrix that contains the known transmitted symbols on its main diagonal, i.e., $ \mathbf{A}_k=\textrm{diag}\Big\{a_k(t_1), a_k(t_2), \cdots,a_k(t_{\bar{N}_{\textrm{DA}}})\Big\}$. Note here that $\widetilde{\mathbf{P}}_k$ is different from $\mathbf{P}_k$ that is defined earlier right after (\ref{projector_manuscript})  as the orthogonal projector over the signal subspace (of the whole antennae array) corresponding to the $k^{th}$ local DA approximation window. 
We associate to $\widetilde{\mathbf{P}}_k$  the operator $\widetilde{\mathbf{P}}_k^{\perp}=\mathbf{I}-\widetilde{\mathbf{P}}_k$ as the projector onto the orthogonal complement of the corresponding signal subspace.\\
Now recall that the estimates of the $i^{th}$ antenna's channel coefficients corresponding to  the $k^{th}$ local DA approximation window, $\widehat{\mathbf{h}}_{i,\textrm{DA}}^{(k)}=[\widehat{h}_{i,\textrm{DA}}^{(k)}(t_1),\widehat{h}_{i,\textrm{DA}}^{(k)}(t_2),\cdots,\widehat{h}_{i,\textrm{DA}}^{(k)}(t_{\bar{N}_{\textrm{DA}}})]^T$, are obtained as:
\begin{eqnarray}\label{local_channel}
\widehat{\mathbf{h}}_{i,\textrm{DA}}^{(k)}=\mathbf{T}\widehat{\mathbf{c}}_{i,\textrm{DA}}^{(k)},
\end{eqnarray}  
where $\widehat{\mathbf{c}}_{i,\textrm{DA}}^{(k)}$ is obtained by extracting the corresponding $k^{th}$ block from $\widehat{\mathbf{c}}_{i,\textrm{DA}}$ obtained in (\ref{Eq.16})  to yield:
\begin{eqnarray}\label{local_polynomial}
\widehat{\mathbf{c}}_{i,\textrm{DA}}^{(k)}=(\bm{\Phi}_k^H\bm{\Phi}_k)^{-1}\bm{\Phi}_k^{H}\widehat{\mathbf{y}}_{i,\textrm{DA}}^{(k)}.
\end{eqnarray}
Therefore, by substituting (\ref{local_polynomial}) back into (\ref{local_channel}), it follows that:
\begin{eqnarray}\label{channel_estimtes}
\widehat{\mathbf{h}}_{i,\textrm{DA}}^{(k)}=\mathbf{T}(\bm{\Phi}_k^H\bm{\Phi}_k)^{-1}\bm{\Phi}_k^{H}\widehat{\mathbf{y}}_{i,\textrm{DA}}^{(k)}.
\end{eqnarray} 
Moreover, from (\ref{local_channel}), we readily see that the $n^{th}$ component, $\widehat{h}_{i,\textrm{DA}}^{(k)}(t_n)$, of $\widehat{\mathbf{h}}_{i,\textrm{DA}}^{(k)}$ is obtained as the inner product between the $n^{th}$ row of $\mathbf{T}$ \Big(i.e., the vector $\mathbf{t}_n=\big[t_n^0,t_n^1,\cdots,t_n^{L-1}\big]^T\Big)$ and $\widehat{\mathbf{c}}_{i,\textrm{DA}}^{(k)}=[\widehat{c}_{i,k}^{(0)},\widehat{c}_{i,k}^{(1)},\cdots,\widehat{c}_{i,k}^{(L-1)}]^T$ leading to:
\begin{eqnarray}\label{elementary_channel}
\widehat{h}_{i,\textrm{DA}}^{(k)}(t_n)=\mathbf{t}_n^T\widehat{\mathbf{c}}_{i,\textrm{DA}}^{(k)}=\sum_{l=0}^{L-1}\widehat{c}_{i,k}^{(l)}t_n^l.
\end{eqnarray}
Now, recall from (\ref{Eq.19}) that the estimated SNR in the DA mode is given:
\begin{equation}\label{estimated_SNR}
\widehat{\rho}_{i,\textrm{DA}}=\frac{\sum_{k=1}^{N/\bar{N}_{\textrm{DA}}}\sum_{n=1}^{\bar{N}_{\textrm{DA}}}\left(\left|a_k(t_n)\right|^2\left |\sum_{l=0}^{L-1}\widehat{c}_{i,k}^{(l)}t_n^l\right|^2\right)}{ N\left(\frac{\bar{N}_{\textrm{DA}}}{N}\sum_{k=1}^{N/\bar{N}_{\textrm{DA}}}2\widehat{\sigma^2}_{k,\textrm{DA}}\right)},
\end{equation}
and owing to (\ref{elementary_channel}), the numerator  of the estimated SNR  in  (\ref{estimated_SNR}) (denoted herafter as ``$Num$'') is expressed as follows:  
\begin{eqnarray}
Num&=&\sum_{k=1}^{N/\bar{N}_{\textrm{DA}}}\sum_{n=1}^{\bar{N}_{\textrm{DA}}}\Bigg(|a_k(t_n)|^2\big|\widehat{h}_{i,\textrm{DA}}^{(k)}(t_n)\big|^2\Bigg).
\end{eqnarray}
By further noticing that:
\begin{eqnarray}
\sum_{n=1}^{\bar{N}_{\textrm{DA}}}\Big(|a_k(t_n)|^2\big|\widehat{h}_{i,\textrm{DA}}^{(k)}(t_n)\big|^2\Big)=\Big|\!\Big|\mathbf{A}_k\widehat{\mathbf{h}}_{i,\textrm{DA}}^{(k)}\Big|\!\Big|^2,
\end{eqnarray}
 it follows that:
\begin{eqnarray}\label{numerator_11}
Num&=&\sum_{k=1}^{N/\bar{N}_{\textrm{DA}}}\Big|\!\Big|\mathbf{A}_k\widehat{\mathbf{h}}_{i,\textrm{DA}}^{(k)}\Big|\!\Big|^2.
\end{eqnarray}
Then, by using  (\ref{channel_estimtes})  and recalling the fact that $\bm{\Phi}_k\triangleq\mathbf{A}_k\mathbf{T}$, we have:
\begin{eqnarray}\label{channel_via_projection}
\mathbf{A}_k\widehat{\mathbf{h}}_{i,\textrm{DA}}^{(k)}=\mathbf{A}_k\mathbf{T}(\bm{\Phi}_k^H\bm{\Phi}_k)^{-1}\bm{\Phi}_k^{H}\mathbf{y}_{i,\textrm{DA}}^{(k)}=\widetilde{\mathbf{P}}_k\mathbf{y}_{i,\textrm{DA}}^{(k)}.
\end{eqnarray}
Then, by substituting (\ref{channel_via_projection}) in (\ref{numerator_11}), it follows that:
\begin{eqnarray}\label{numerator_1}
Num=\sum_{k=1}^{N/\bar{N}_{\textrm{DA}}}\Big|\!\Big|\widetilde{\mathbf{P}}_k\mathbf{y}_{i,\textrm{DA}}^{(k)}\Big|\!\Big|^2.
\end{eqnarray}
On the other hand, the denominator of the SNR estimate in (\ref{estimated_SNR}) is given by:
\begin{eqnarray}\label{Denom}
Denom=N\left(\frac{\bar{N}_{\textrm{DA}}}{N}\sum_{k=1}^{N/\bar{N}_{\textrm{DA}}}2\widehat{\sigma^2}_{k,\textrm{DA}}\right),
\end{eqnarray}
and since $2\widehat{\sigma^2}_{k,\textrm{DA}}$ is obtained from  (\ref{projector_manuscript}) as:
\begin{eqnarray}\label{local_noise_variance_estimate}
2\widehat{\sigma^2}_{k,\textrm{DA}}&=&\frac{1}{{\bar{N}_{\textrm{DA}}}N_r}\left[{{\bf y}^{(k)}_{\textrm{DA}}}^H{\bf P}_k^{\perp}{\bf y}^{(k)}_{\textrm{DA}}\right],
\end{eqnarray}
with ${\bf P}_k\triangleq\textrm{Blkdiag}\Big\{\widetilde{\mathbf{P}}_1,\widetilde{\mathbf{P}}_2,\cdots,\widetilde{\mathbf{P}}_{N/\bar{N}_{\textrm{DA}}}\Big\}$, we obtain by substituting (\ref{local_noise_variance_estimate}) back in (\ref{Denom}) the following result:
\begin{eqnarray}\label{Denom_1}
Denom=\frac{1}{N_r}\sum_{k=1}^{N/\bar{N}_{\textrm{DA}}}{{\bf y}^{(k)}_{\textrm{DA}}}^H{\bf P}_k^{\perp}{\bf y}^{(k)}_{\textrm{DA}}=\frac{1}{N_r}\sum_{k=1}^{N/\bar{N}_{\textrm{DA}}}\Big|\!\Big|{\bf P}_k^{\perp}{\bf y}^{(k)}_{\textrm{DA}}\Big|\!\Big|^2.
\end{eqnarray}
Raclling that ${\bf y}^{(k)}_{\textrm{DA}}\triangleq\Big[{\bf y}^{(1)^T}_{i,\textrm{DA}}~{\bf y}^{(2)^T}_{i,\textrm{DA}}\cdots{\bf y}^{(N_r)^T}_{i,\textrm{DA}}\Big]^T$, it can be easily shown that:
\begin{eqnarray}\label{Denom_2}
Denom&=&\frac{1}{N_r}\sum_{k=1}^{N/\bar{N}_{\textrm{DA}}}\sum_{i=1}^{N_r}{{\bf y}^{(k)^H}_{i,\textrm{DA}}}\widetilde{\bf P}_k^{\perp}{\bf y}^{(k)}_{i,\textrm{DA}}\nonumber\\
&=&\frac{1}{N_r}\sum_{i=1}^{N_r}\sum_{k=1}^{N/\bar{N}_{\textrm{DA}}}{{\bf y}^{(k)^H}_{i,\textrm{DA}}}\widetilde{\bf P}_k^{\perp}{\bf y}^{(k)}_{i,\textrm{DA}}.
\end{eqnarray}
Now, let $\mathbf{y}_{i,\textrm{DA}}\triangleq\Big[\mathbf{y}_{i,\textrm{DA}}^{(1)^T}~{\mathbf{y}_{i,\textrm{DA}}^{(2)^T}}\cdots\mathbf{y}_{i,\textrm{DA}}^{(N/\bar{N}_{\textrm{DA}})^T}\Big]^T$ be a vector that contains all the received samples over the $i^{th}$ antenna elements. Thus, the SNR estimate at the $i^{th}$ antenna element is obtained from (\ref{numerator_1}) and (\ref{Denom_2}) as follows:
\begin{equation}\label{estimated_SNR_projections}
\widehat{\rho}_{i,\textrm{DA}}=\frac{Num}{Denom}=\frac{\mathbf{y}^H_{i,\textrm{DA}}\widetilde{\mathbf{P}}\mathbf{y}_{i,\textrm{DA}}}{\frac{1}{N_r}\sum_{i=1}^{N_r}\sum_{k=1}^{N/\bar{N}_{\textrm{DA}}}{{\bf y}^{(k)^H}_{i,\textrm{DA}}}\widetilde{\bf P}_k^{\perp}{\bf y}^{(k)}_{i,\textrm{DA}}}.
\end{equation}
Next, in order to find the distribution of $\widehat{\rho}_{i,\textrm{DA}}$, we will first proceed to finding the distributions of $Num$ and $Denom$ separately. To that end,
recall first from  (\ref{Eq.12}) that (when $N_p=1$):
\begin{eqnarray}
{\bf y}^{(k)}_{i,\textrm{DA}}=\bm{\Phi}_k\mathbf{c}_{i,k}+\mathbf{w}_{i,k},
\end{eqnarray}
where $\mathbf{w}_{i,k}\sim\mathcal{N}\big(\mathbf{0},2\sigma^2\mathbf{I}_{\bar{N}_{\textrm{DA}}}\big)$ with $\mathbf{I}_{\bar{N}_{\textrm{DA}}}$ being the $\bar{N}_{\textrm{DA}}\times \bar{N}_{\textrm{DA}}$ identity matrix. Therefore, the mean and covariance matrix of ${\bf y}^{(k)}_{i,\textrm{DA}}$ are given by:
\begin{eqnarray}
\mathbf{m}_{i,k}&=&\bm{\Phi}_k\mathbf{c}_{i,k},\\
\mathbf{R}_{{\bf y}^{(k)}_{i,\textrm{DA}}{\bf y}^{(k)}_{i,\textrm{DA}}}&=&2\sigma^2\mathbf{I}_{\bar{N}_{\textrm{DA}}}.
\end{eqnarray}
Therefore, if we define the following transformed random vector:
\begin{eqnarray}\label{transformation}
\widetilde{\mathbf{y}}_{i,\textrm{DA}}^{(k)}\triangleq\frac{1}{\sqrt{2\sigma^2}}\mathbf{y}_{i,\textrm{DA}}^{(k)},
\end{eqnarray}
then we immediately have $\widetilde{\mathbf{y}}_{i,\textrm{DA}}^{(k)}\sim\mathcal{N}\big(\frac{1}{\sqrt{2\sigma^2}}\mathbf{m}_{i,k},\mathbf{I}_{\bar{N}_{\textrm{DA}}}\big)$. Now, since  $\widetilde{\mathbf{P}}_k$ is a Hermetian matrix then it can be diagonalized as follows:
\begin{eqnarray}\label{eigenvalue}
\widetilde{\mathbf{P}}_k=\mathbf{U}_k\mathbf{D}\mathbf{U}_k^H,
\end{eqnarray}
where $\mathbf{U}_k$ is a \textit{unitary} matrix (i.e., $\mathbf{U}_k\mathbf{U}_k^H=\mathbf{U}_k^H\mathbf{U}_k=\mathbf{I}_{\bar{N}_{\textrm{DA}}}$) and $\mathbf{D}=\textrm{diag}\big\{\lambda_1,\lambda_2,\cdots,\lambda_{\bar{N}_{\textrm{DA}}}\big\}$ is a diagonal matrix that contains the $\bar{N}_{\textrm{DA}}$ eigenvalues of $\widetilde{\mathbf{P}}_k$ (which are all  \textit{positive}). Moreover, since $\widetilde{\mathbf{P}}_k\widetilde{\mathbf{P}}_k^H=\widetilde{\mathbf{P}}_k$, we have:
\begin{eqnarray}
\mathbf{U}_k\mathbf{D}^2\mathbf{U}_k^H=\mathbf{U}_k\mathbf{D}\mathbf{U}_k^H,
\end{eqnarray} 
which means that $\mathbf{D}^2=\mathbf{D}$ or equivalently $\lambda_n^2=\lambda_n$ for $n=1,2,\cdots\bar{N}_{\textrm{DA}}$ and, therefore, \Big\{$\lambda_n=0$ or $\lambda_n=1$ for $n=1,2,\cdots\bar{N}_{\textrm{DA}}$\Big\}. However, since $\mathbf{T}$ is a Vondermende matrix, it is of (full) rank $L$ and since $\mathbf{A}_k$ is a diagonal matrix, it follows that $\mathbf{\Phi}_k$ is also of rank $L$. Consequently, the projection matrix  $\widetilde{\mathbf{P}}_k=\mathbf{\Phi}_k(\mathbf{\Phi}_k^H\mathbf{\Phi}_k)^{-1}\mathbf{\Phi}_k^H$ is also of rank $L$ and, therefore, we have exactly $L$ eigenvalues that are equal to one and the others are exactly zero. In the following we assume (without loss of generality) that the first $L$ eigenvalues are non-zero. That is $\lambda_n=1$ for $n=1,2,\cdots,L$ and $\lambda_n=0$ for $n=L+1,L+2,\cdots,\bar{N}_{\textrm{DA}}$, which means:
\begin{eqnarray}\label{matrix_D}
\mathbf{D}=\textrm{diag}\big\{\underbrace{1,1\cdots,1}_{L~\textrm{times}},0,0,\cdots,0\big\}~~~~(\bar{N}_{\textrm{DA}}\times\bar{N}_{\textrm{DA}}~\textrm{matrix})\nonumber\\
\end{eqnarray}   
Now, combining (\ref{transformation}) and (\ref{eigenvalue}) and using the fact that $\mathbf{U}_k$ is a \textit{unitary} matrix, it follows that: 
\begin{eqnarray}
\Big|\!\Big|\widetilde{\mathbf{P}}_k\mathbf{y}_{i,\textrm{DA}}^{(k)}\Big|\!\Big|^2=\frac{1}{2\sigma^2}\Big|\!\Big|\mathbf{U}_k\mathbf{D}\mathbf{U}_k^H\widetilde{\mathbf{y}}_{i,\textrm{DA}}^{(k)}\Big|\!\Big|^2
=\frac{1}{2\sigma^2}\Big|\!\Big|\mathbf{D}\mathbf{U}_k^H\widetilde{\mathbf{y}}_{i,\textrm{DA}}^{(k)}\Big|\!\Big|^2.
\end{eqnarray}
By further defining the transformed received vector:
\begin{eqnarray}
\bar{\mathbf{z}}^{(k)}_{i,\textrm{DA}}\triangleq\mathbf{U}_k^H\widetilde{\mathbf{y}}_{i,\textrm{DA}}^{(k)},
\end{eqnarray}
and again using the fact that $\mathbf{U}_k$ is a \textit{unitary} matrix, it follows that $\bar{\mathbf{z}}^{(k)}_{i,\textrm{DA}}\sim\mathcal{N}\big(\frac{1}{\sqrt{2\sigma^2}}\mathbf{U}_k^H\mathbf{m}_{i,k},\mathbf{I}_{\bar{N}_{\textrm{DA}}}\big)$ and:
\begin{eqnarray}\label{projection_expanded}
\Big|\!\Big|\widetilde{\mathbf{P}}_k\mathbf{y}_{i,\textrm{DA}}^{(k)}\Big|\!\Big|^2=\frac{1}{2\sigma^2}\Big|\!\Big|\mathbf{D}\bar{\mathbf{z}}^{(k)}_{i,\textrm{DA}}\Big|\!\Big|^2=\frac{1}{2\sigma^2}\sum_{l=1}^{L}\left[\bar{\mathbf{z}}^{(k)}_{i,\textrm{DA}}\right]_l^2,
\end{eqnarray}
in which $\left[\bar{\mathbf{z}}^{(k)}_{i,\textrm{DA}}\right]_l$ is used to denote the $l^{th}$ element of the vector $\bar{\mathbf{z}}^{(k)}_{i,\textrm{DA}}$ and where the last equality follows from the fact that only the first diagonal entries of $\mathbf{D}$ are non-zero and are all equal to one [see (\ref{matrix_D})]. By plugging (\ref{projection_expanded}) back into (\ref{numerator_1}), we obtain:
\begin{eqnarray}\label{numerator_expanded}
Num=\frac{1}{2\sigma^2}\sum_{k=1}^{N/\bar{N}_{\textrm{DA}}}\sum_{l=1}^{L}\left[\bar{\mathbf{z}}^{(k)}_{i,\textrm{DA}}\right]_l^2.
\end{eqnarray} 
In addition, since the vector $\bar{\mathbf{z}}^{(k)}_{i,\textrm{DA}}$ is Gaussian distributed according to  $\bar{\mathbf{z}}^{(k)}_{i,\textrm{DA}}\sim\mathcal{N}\big(\frac{1}{\sqrt{2\sigma^2}}\mathbf{U}_k^H\mathbf{m}_{i,k},\mathbf{I}_{\bar{N}_{\textrm{DA}}}\big)$, then its elements $\left[\bar{\mathbf{z}}^{(k)}_{i,\textrm{DA}}\right]_l$ are independent and Gaussian distributed according to:
\begin{eqnarray}
\left[\bar{\mathbf{z}}^{(k)}_{i,\textrm{DA}}\right]_l\sim\mathcal{N}\left(\textstyle\frac{1}{\sqrt{2\sigma^2}}[\mathbf{U}_k]_{:,l}^H\mathbf{m}_{i,k},1\right),
\end{eqnarray}
where $[\mathbf{U}_k]_{:,l}$ is used to denote the $l^{th}$ column of the matrix $\mathbf{U}_k$. Consequently, $2\sigma^2\times Num$ is a sum of the squares of $NL/\bar{N}_{\textrm{DA}}$ \textit{independent} Gaussian random variables all having unit variance but non-zero means and, therefore,  is  chi-square distributed with $\nu_1=\frac{N}{\bar{N}_{\textrm{DA}}}L$ \textit{degrees of freedom}  and \textit{noncentrality parameter}:
\begin{eqnarray}
\lambda&=&\sum_{k=1}^{N/\bar{N}_{\textrm{DA}}}\sum_{l=1}^{L}\Big|\textstyle\frac{1}{\sqrt{2\sigma^2}}[\mathbf{U}_k]_{:,l}^H\mathbf{m}_{i,k}\Big|^2,\nonumber\\
&=&\frac{1}{2\sigma^2}\sum_{k=1}^{N/\bar{N}_{\textrm{DA}}}\Big|\!\Big|\mathbf{D}\mathbf{U}_k^H\mathbf{m}_{i,k}\Big|\!\Big|^2
\end{eqnarray} 
where the last quality follows from the fact that the first $L$ diagonal entries of $\mathbf{D}$ are equal to one and the remaining $\bar{N}_{\textrm{DA}}-L$ diagonal ones are all equal to zero. Furthermore, by recalling that $\mathbf{m}_{i,k}=\bm{\Phi}_k\mathbf{c}_{i,k}$ and that $\mathbf{D}^2=\mathbf{D}$, it follows that: 
\begin{eqnarray}
\lambda&=&\frac{1}{2\sigma^2}\sum_{k=1}^{N/\bar{N}_{\textrm{DA}}}\Big|\!\Big|\mathbf{D}\mathbf{U}_k^H\mathbf{m}_{i,k}\Big|\!\Big|^2\nonumber\\
&=&\frac{1}{2\sigma^2}\sum_{k=1}^{N/\bar{N}_{\textrm{DA}}}\mathbf{m}_{i,k}^H\mathbf{U}_k\mathbf{D}\mathbf{U}_k^H\mathbf{m}_{i,k}\nonumber\\
&=&\frac{1}{2\sigma^2}\sum_{k=1}^{N/\bar{N}_{\textrm{DA}}}\mathbf{m}_{i,k}^H\mathbf{P}_k\mathbf{m}_{i,k}\nonumber\\
&=&\frac{1}{2\sigma^2}\sum_{k=1}^{N/\bar{N}_{\textrm{DA}}}\mathbf{c}_{i,k}^H\bm{\Phi}_k^H\bm{\Phi}_k(\bm{\Phi}_k^H\bm{\Phi}_k)^{-1}\bm{\Phi}_k^H\bm{\Phi}_k^H\mathbf{c}_{i,k}\nonumber\\
&=&\frac{1}{2\sigma^2}\sum_{k=1}^{N/\bar{N}_{\textrm{DA}}}\mathbf{c}_{i,k}^H\bm{\Phi}_k^H\bm{\Phi}_k\mathbf{c}_{i,k}\nonumber\\
&=&\frac{1}{2\sigma^2}\sum_{k=1}^{N/\bar{N}_{\textrm{DA}}}\Big|\!\Big|\bm{\Phi}_k\mathbf{c}_{i,k}\Big|\!\Big|^2\nonumber\\
&=&\frac{1}{2\sigma^2}\sum_{k=1}^{N/\bar{N}_{\textrm{DA}}}\big|\!\big|\mathbf{h}_{i,k}\big|\!\big|^2\nonumber\\
&=&\frac{\big|\!\big|\mathbf{h}_{i}\big|\!\big|^2}{2\sigma^2}\nonumber\\
&=&N\rho_i
\end{eqnarray} 
In conclusion, we have $2\sigma^2\times Num$ is a \textit{noncentral} chi-distributed, i.e.:
\begin{eqnarray}
2\sigma^2 Num\sim\chi^2_{\nu_1}(\lambda),
\end{eqnarray}
 with  $\nu_1=\frac{N}{\bar{N}_{\textrm{DA}}}L$ \textit{degrees of freedom}  and \textit{noncentrality parameter} $\lambda=N\rho_i$.\\
Now recall from (\ref{Denom_2}) that the denominator is equal to:
\begin{eqnarray}\label{Denom_2_2}
Denom&=&\frac{1}{N_r}\sum_{i=1}^{N_r}\sum_{k=1}^{N/\bar{N}_{\textrm{DA}}}{{\bf y}^{(k)^H}_{i,\textrm{DA}}}\widetilde{\bf P}_k^{\perp}{\bf y}^{(k)}_{i,\textrm{DA}}.
\end{eqnarray}
Similarly, by noticing that $\mathbf{P}^{\perp}_k$ is of rank  $\bar{N}_{\textrm{DA}}-L$ and recurring to equivalent manipulations, it can be shown that the denominator can be rewritten in the following form:
\begin{eqnarray}
Denom&=&\frac{1}{2\sigma^2N_r}\sum_{i=1}^{N_r}\sum_{k=1}^{N/\bar{N}_{\textrm{DA}}}\sum_{l=1}^{\bar{N}_{\textrm{DA}-L}}\left[\bar{\mathbf{v}}^{(k)}_{i,\textrm{DA}}\right]_l^2,
\end{eqnarray}
where $\left[\bar{\mathbf{v}}^{(k)}_{i,\textrm{DA}}\right]_l^2$ are the components of another transformed observation vector which are Gaussian distributed with zero mean and unit variance. Hence, the random variable $2\sigma^2N_r\times Denom$ follows a \textit{central} chi-distribution \cite{ref20}, i.e.:
\begin{eqnarray}
2\sigma^2N_rDenom\sim\chi^2_{\nu_2},
\end{eqnarray}
with $\nu_2=N_r\frac{N}{\bar{N}_{\textrm{DA}}}(\bar{N}_{\textrm{DA}}-L)=N_r(N-\frac{N}{\bar{N}_{\textrm{DA}}}L)$ degrees of freedom. Moreover, $Num$ and $Denom$ involve projection onto a signal subspace and its orthogonal complement, respectively, and hence the two chi-distributed random variables are independent. In conclusion, we have: 
\begin{eqnarray}
\widehat{\rho}_{i,\textrm{DA}}=\frac{\frac{1}{2\sigma^2}\chi^2_{\nu_1}(\lambda)}{\frac{1}{2\sigma^2N_r}\chi^2_{\nu_2}},
\end{eqnarray} 
which implies that the scaled estimated SNR over each $i^{th}$ antenna element verfies:
\begin{eqnarray}
\frac{\nu_2}{\nu_1}\frac{1}{N_r}\widehat{\rho}_{i,\textrm{DA}}=\frac{(N-\frac{N}{\bar{N}_{\textrm{DA}}}L)}{\frac{N}{\bar{N}_{\textrm{DA}}}L}\widehat{\rho}_{i,\textrm{DA}}=\frac{\chi^2_{\nu_1}(\lambda)/\nu_1}{\chi^2_{\nu_2}/\nu_2}=F_{\nu_1,\nu_2}(\lambda),
\end{eqnarray} 
where $F_{\nu_1,\nu_2}(\lambda)$ is a \textit{noncentral} $F$ distribution with a \textit{noncentrality parameter} $\lambda=N\rho_i$ and degrees of freedom  $\nu_1=\frac{N}{\bar{N}_{\textrm{DA}}}L$ and  $\nu_2=N_r(N-\frac{N}{\bar{N}_{\textrm{DA}}}L)$.

\section*{Appendix B\\Details about the derivation of the bias and the variance}
\noindent By using $\epsilon= L/\bar{N}_{\textrm{DA}}$, it follows immediately from (\ref{Eq.20}) that:
\begin{eqnarray}\label{Eq.23_appendix_A}
\textrm{E}\{\widehat{\rho}_{i,\textrm{DA}}\}=\textrm{E}\left\{\frac{\epsilon}{2(1-\epsilon)}F\right\}= \frac{\epsilon}{2(1-\epsilon)}\textrm{E}\left\{F\right\}.
\end{eqnarray}
Moreover, by substituting $\lambda=2N\rho_i$,  $v_1=N\epsilon$ and $v_2=2N_rN(1-\epsilon)$ in (\ref{Eq.21}), it follows that:
\begin{equation}\label{Eq.21_Appendix_A}
\textrm{E}\{F\}=\frac{2N_rN(1-\epsilon)(N\epsilon+2N\rho_i)}{N\epsilon\big[2N_rN(1-\epsilon)-2\big]}.
\end{equation} 
Then, by recognizing some easy simplifications, one obtains:
\begin{equation}\label{Eq.21_Appendix_A_simplified}
\textrm{E}\{F\}=\frac{1-\epsilon}{\epsilon}\left(\frac{N_rN(2\rho_i+\epsilon)}{N_rN(1-\epsilon)-1}\right).
\end{equation} 
Therefore, by using (\ref{Eq.21_Appendix_A_simplified}) in (\ref{Eq.23_appendix_A}), it follows that:
\begin{eqnarray}\label{Eq.23_appendix_A_simplified}
\textrm{E}\{\widehat{\rho}_{i,\textrm{DA}}\}= \displaystyle\frac{N_rN}{N_rN(1-\epsilon)-1}\left(\rho_i+\frac{\epsilon}{2}\right).
\end{eqnarray}
Now, (\ref{Eq.24}) is obtained in the same way, i.e., first by substituting $\lambda=2N\rho_i$,  $v_1=N\epsilon$ and $v_2=2N_rN(1-\epsilon)$ in (\ref{Eq.22}) (with  some easy simplifications) and then injecting the result in the following identity: 
\begin{eqnarray}\label{variance_appendix_A}
\textrm{Var}\{\widehat{\rho}_{i,\textrm{DA}}\}=\textrm{Var}\left\{\frac{\epsilon}{2(1-\epsilon)}F\right\}= \left(\frac{\epsilon}{2(1-\epsilon)}\right)^2\textrm{Var}\left\{F\right\}.
\end{eqnarray}
The exact bias of $\widehat{\rho}_{i,\textrm{DA}}$,  which is given by $\textrm{Bias}\{\widehat{\rho}_{i,\textrm{DA}}\}=\textrm{E}\{\widehat{\rho}_{i,\textrm{DA}}-\rho_i\}=\textrm{E}\{\widehat{\rho}_{i,\textrm{DA}}\}-\rho_i$, is then easily obtained from (\ref{Eq.23_appendix_A_simplified})  as given by (\ref{Eq.25}). Furthermore, it follows from  (\ref{Eq.23_appendix_A_simplified}) that:
\begin{eqnarray}\label{Eq.26_Appendix}
\textrm{E}\!\left\{\!\frac{N_rN(1-\epsilon)-1}{N_rN}\widehat{\rho}_{i,\textrm{DA}}\!-\frac{\epsilon}{2}\!\right\}&\!\!\!\!\!=\!\!\!\!\!&\frac{N_rN(1-\epsilon)-1}{N_rN}\textrm{E}\left\{\widehat{\rho}_{i,\textrm{DA}}\right\}-\frac{\epsilon}{2}\nonumber\\
&\!\!\!\!\!=\!\!\!\!\!&\rho_i,
\end{eqnarray}
which simply implies that:
\begin{equation}\label{unbiased_appendix}
 \widehat{\rho}_{i,\textrm{DA}}^{~\textrm{UB}}=\frac{N_rN(1-\epsilon)-1}{N_rN}\widehat{\rho}_{i,\textrm{DA}}-\frac{\epsilon}{2},
\end{equation}
 is indeed an \textit{unbiased} estimator of the per-antenna SNRs. By using the identity $\textrm{Var}\{aX+b\}=a^2\textrm{Var}\{X\}$ for any random variable $X$ and any real number $a$, the variance of $\widehat{\rho}_{i,\textrm{DA}}^{~\textrm{UB}}$ is given by:
\begin{eqnarray}
\textrm{Var}\{\widehat{\rho}_{i,\textrm{DA}}^{~\textrm{UB}}\}=\left(\frac{N_rN(1-\epsilon)-1}{N_rN}\right)^2\textrm{Var}\{\widehat{\rho}_{i,\textrm{DA}}\},
\end{eqnarray} 
which is further simplified  using (\ref{Eq.24}) in order to obtain the result in (\ref{Eq.28}).
\section*{Appendix C\\Derivation of the DA CRLB}
To assess the performance of the new unbiased DA ML estimator, we need to compare its variance to a theoretical lower bound. Thus, we derive in this Appendix the corresponding DA CRLB. Here, for some reasons that are better clarified in sections IV and V, we are interested in comparing our estimators against the lowest possible bound (i.e., the best achievable performance). Without loss of generality, we hence consider an ideal scenario where all the transmitted symbols are assumed to be perfectly known (i.e., $N_p=1$ or equivalently $N'=N$). Now, we define the following parameter vector:
\begin{eqnarray}\label{Eq.30}
{\bm\theta}'=[{\bm \alpha}^T, {\bm \beta}^T, \sigma^2]^T,
\end{eqnarray}
where $\bm\alpha=\Re\{\bf h\}$ and $\bm\beta=\Im\{\bf h\}$ denote the real and imaginary parts of the vector ${\bf h}=[{\bf h}_1^T,{\bf h}_2^T,\cdots,{\bf h}_{N_r}^T]^T$  that contains the true channel coefficients over all the receiving antenna elements and the entire observation window. The CRLB for the DA SNR estimation over the $i^{th}$ antenna is given by:
\begin{eqnarray}\label{Eq.31}
\textrm{CRLB}_{\textrm{DA}}(\rho_i)=\bigg(\frac{\partial\rho_i}{\partial\bm\theta '}\bigg)^T{\bf I}^{-1}_{\textrm{DA}}(\bm\theta ')\bigg(\frac{\partial\rho_i}{\partial\bm\theta '}\bigg),
\end{eqnarray}
where $\rho_i=({\bf Ah}_i)^{H}{\bf Ah}_i/N(2\sigma^2)$ with ${\bf{A}}$ being a diagonal matrix containing the $N'=N$ transmitted pilot symbols and where ${\bf I}_{\textrm{DA}}(\bm\theta ')$ denotes the Fisher information matrix (FIM) whose entries are defined as:
\begin{eqnarray}\label{Eq.32}
\big[{\bf I}_{\textrm{DA}}(\bm{\theta} ')\big]_{i,l}=-\textrm{E}_{{\bf y}_{\textrm{DA}}}\bigg\{\frac{\partial ^2\ln\big(p({\bf y}_{\textrm{DA}};\bm\theta ')\big)}{\partial\bm\theta_i'\partial\bm\theta_l'^T}\bigg\},
\end{eqnarray}
where
\begin{eqnarray}\label{Eq.33}
p({\bf y}_{\textrm{DA}};\bm\theta ')&\!\!\!\!\!=\!\!\!\!\!&\frac{\exp\!\bigg\{\!\!-\displaystyle\frac{1}{2\sigma^2}\displaystyle\sum_{i=1}^{N_r}({\bf y}_{i,\textrm{DA}}\!-\!{\bf A} {\bf h}_i)^{H}({\bf y}_{i,\textrm{DA}}\!-\!{\bf A} {\bf h}_i)\!\bigg\}}{(2\pi\sigma^2)^{NN_r}}.\nonumber\\
&\!\!\!\!\!\!\!\!\!\!&\!\!\!\!\!\!\!\!\!
\end{eqnarray}
In (\ref{Eq.33}), ${\bf y}_{i,\textrm{DA}}$ ans ${\bf y}_{\textrm{DA}}$ are given by:
\begin{eqnarray}
 {\bf y}_{i,\textrm{DA}}&=&[y_{i,\textrm{DA}}(t_1), y_{i,\textrm{DA}}(t_2),\cdots,y_{i,\textrm{DA}}(t_N)]^T,\nonumber\\
 {\bf y}_{\textrm{DA}}&=&[{\bf y}_{1,\textrm{DA}}^T,{\bf y}_{2,\textrm{DA}}^T,\cdots,{\bf y}_{N_r,\textrm{DA}}^T]^T,\nonumber
\end{eqnarray}
with $t_n=nT_s$ for $n=1,2,\cdots,N$. Starting from (\ref{Eq.33}), we will now derive the analytical expression for the FIM.
In fact, by recalling that ${\bf h}_i={\bm\alpha}_i+j{\bm\beta}_i$ where ${\bm\alpha}_i$ and ${\bm\beta}_i$ stand for the real and imaginary parts of ${\bf h}_i$, respectively, we can obtain the required partial derivatives in (\ref{Eq.32}) as follows:
\begin{eqnarray}\label{Eq.61}
\frac{\partial^2\ln(P({\bf y}_{\textrm{DA}};{\bm\theta}'))}{\partial{\bm\alpha}_i\partial{\bm\alpha}_i^T}=\frac{\partial^2\ln(P({\bf y}_{\textrm{DA}};{\bm\theta}'))}{\partial{\bm\beta}_i\partial{\bm\beta}_i^T}=-\frac{1}{\sigma^2}{\bf A}^H{\bf A},
\end{eqnarray}
\begin{eqnarray}\label{Eq.62}
\frac{\partial^2\ln(P({\bf y}_{\textrm{DA}};{\bm\theta}'))}{\partial{\sigma^2}\partial{\bm\alpha}_i^T}=\frac{1}{2\sigma^4}\big(2{\bm\alpha}_i^T{\bf A}^H{\bf A}-2\Re\{{\bf y}_{i,{\textrm{DA}}}^{H}{\bf A}\}\big),
\end{eqnarray}
\begin{eqnarray}\label{Eq.63}
\frac{\partial^2\ln(P({\bf y}_{\textrm{DA}};{\bm\theta}'))}{\partial{\sigma^2}\partial{\bm\beta}_i^T}=\frac{1}{2\sigma^4}\big(2{\bm\beta}_i^T{\bf A}^H{\bf A}-2\Im\{{\bf y}_{i,{\textrm{DA}}}^{H}{\bf A}\}\big),
\end{eqnarray}
and
\begin{eqnarray}\label{Eq.64}
\!\!\!\!\frac{\partial^2\ln(P({\bf y}_{\textrm{DA}};{\bm\theta}'))}{\partial{\sigma^2}^2}&\!\!\!\!=\!\!\!\!&\frac{NN_r}{\sigma^4}-\nonumber\\
\!\!\!\!&\!\!\!\!\!\!\!\!&\!\!\!\!\!\!\!\!\!\!\!\!\!\!\!\!\frac{1}{\sigma^6}\sum_{i=1}^{N_r}({\bf y}_{i,{\textrm{DA}}}-{\bf A}{\bf h}_i)^{H}({\bf y}_{i,{\textrm{DA}}}-{\bf A}{\bf h}_i).
\end{eqnarray}
Moreover, it is easy to verify that:
\begin{eqnarray}\label{Eq.65}
\frac{\partial^2\ln(P({\bf y}_{\textrm{DA}};{\bm\theta}'))}{\partial{\bm\beta}_i\partial{\bm\alpha}_i^T}&=&\frac{\partial^2\ln(P({\bf y}_{\textrm{DA}};{\bm\theta}'))}{\partial{\bm\alpha}_i\partial{\bm\alpha}_l^T}\nonumber\\
&=&\frac{\partial^2\ln(P({\bf y}_{\textrm{DA}};{\bm\theta}'))}{\partial{\bm\beta}_i\partial{\bm\beta}_l^T}={\bf 0}_{N},
\end{eqnarray}
for $1\leq i\leq N_r$ and $1\leq l\leq N_r$  with $i\neq l$. 
Additionally, the expected values of the previously derived partial derivatives with respect to ${\bf y}_{\textrm{DA}}$ are given by:
\begin{eqnarray}\label{Eq.66}
\textrm{E}_{{\bf y}_{\textrm{DA}}}\bigg\{\frac{\partial ^2\ln\big(P({\bf y}_{\textrm{DA}};\bm\theta ')\big)}{\partial\bm\alpha_i\partial\bm\alpha_i^T}\bigg\}&=&\textrm{E}_{{\bf y}_{\textrm{DA}}}\bigg\{\frac{\partial ^2\ln\big(P({\bf y}_{\textrm{DA}};\bm\theta ')\big)}{\partial\bm\beta_i\partial\bm\beta_i^T}\bigg\}\nonumber\\
&=&-\frac{1}{\sigma^2}{\bf A}^H{\bf A}\\
\label{Eq.67}
\textrm{E}_{{\bf y}_{\textrm{DA}}}\bigg\{\frac{\partial^2\ln(P({\bf y}_{\textrm{DA}};{\bm\theta}'))}{\partial{\sigma^2}^2}\bigg\}&=&-\frac{NN_r}{\sigma^4},
\end{eqnarray}
And it can be easily shown that:
\begin{eqnarray}\label{Eq.68}
\textrm{E}_{{\bf y}_{\textrm{DA}}}\bigg\{\frac{\partial ^2\ln\big(P({\bf y}_{\textrm{DA}};\bm\theta ')\big)}{\partial\sigma^2\partial\bm\alpha_i^T}\bigg\}&=&\textrm{E}_{{\bf y}_{\textrm{DA}}}\bigg\{\frac{\partial ^2\ln\big(P({\bf y}_{\textrm{DA}};\bm\theta ')\big)}{\partial\sigma^2\partial\bm\beta_i^T}\bigg\}\nonumber\\
&=&{\bf 0}_{1\times N}.
\end{eqnarray}
Now using:
\begin{eqnarray}\label{Eq.69}
\big[{\bf I}_{\textrm{DA}}(\bm\theta ')\big]_{i,l}=-\textrm{E}_{{\bf y}_{\textrm{DA}}}\bigg\{\frac{\partial ^2\ln\big(P({\bf y}_{\textrm{DA}};\bm\theta ')\big)}{\partial\bm\theta_i'\partial\bm\theta_l'^T}\bigg\},
\end{eqnarray}
we can finally derive the analytical expression for the FIM as follows:
\begin{eqnarray}\label{Eq.70}
{\bf I}_{\textrm{DA}}(\bm\theta ')=\begin{pmatrix} \frac{{\bf A}^H{\bf A}}{\sigma^2}& {\bf 0}_{N} & \cdots & {\bf 0}_{N} & {\bf 0}_{N\times 1}\\ {\bf 0}_{N} &\ddots & \ddots & \vdots & \vdots\\ \vdots & \ddots & \ddots & {\bf 0}_{N} & {\bf 0}_{N\times 1}\\ {\bf 0}_{N} & \cdots & {\bf 0}_{N} & \frac{{\bf A}^H{\bf A}}{\sigma^2} & {\bf 0}_{N\times 1}\\
{\bf 0}_{1\times N} & \cdots & \cdots & {\bf 0}_{1\times N} & \frac{NN_r}{\sigma^4}\end{pmatrix},
\end{eqnarray}
which turns out to be a block-diagonal matrix whose inverse is straightforward. Moreover, by recalling that $\rho_i=({\bf Ah}_i)^{H}{\bf Ah}_i/N(2\sigma^2)$ and ${\bf h}_i={\bm\alpha}_i+j{\bm\beta}_i$, it is easy to verify that:
\begin{eqnarray}\label{Eq.34}
\rho_i=\frac{\bm\alpha_i^T{\bf A}^H{\bf A}\bm\alpha_i+\bm\beta_i^T{\bf A}^H{\bf A}\bm\beta_i}{N(2\sigma^2)},
\end{eqnarray}
from which it can be shown that \cite{cookbook}:
\begin{eqnarray}\label{Eq.35}
\!\!\!\!\!\!\!\!\!\!\!\!&&\frac{\partial\rho_i}{\partial\bm\alpha_i}=\frac{{\bf A}^H{\bf A}\bm\alpha_i}{N\sigma^2},~\frac{\partial\rho_i}{\partial\bm\beta_i}=\frac{{\bf A}^H{\bf A}\bm\beta_i}{N\sigma^2},~\frac{\partial\rho_i}{\partial\sigma^2}=\frac{-({\bf Ah}_i)^{H}{\bf Ah}_i}{2N\sigma^4},\nonumber\\
\!\!\!\!\!\!\!\!\!\!\!\!&&
\end{eqnarray}
and ${\partial\rho_i}/{\partial\bm\alpha_l}={\partial\rho_i}/{\partial\bm\beta_l}={\bf 0}_{1\times N}~~ \textrm{for}~i\neq l$. Finally, by using this result, injecting (\ref{Eq.70})-(\ref{Eq.35}) in (\ref{Eq.31}) and after some algebraic manipulations, a simple closed-form expression for the CRLB of the DA \textit{instantaneous} SNR estimates is obtained as follows:
\begin{eqnarray}\label{Eq.36_Appendix}
\textrm{CRLB}_\textrm{DA}(\rho_i)=\frac{\rho_i}{N}\left(2+\frac{\rho_i}{N_r}\right).
\end{eqnarray}

\end{document}